\documentclass[pra,twocolumn]{revtex4-1}
\usepackage[pdftex]{graphicx}
\usepackage[pdftex]{epsfig}
\usepackage{epstopdf}
\usepackage[caption=false]{subfig}
\usepackage{color}
\usepackage{amssymb,amsmath}
\usepackage{float}
\usepackage{dcolumn}
\usepackage{multirow}
\usepackage{hyperref}
\usepackage{bm}
\hypersetup{colorlinks=true linkcolor={blue} }
\def\mbb{{\mathbf b}}
\def\mbB{{\mathbf B}}

\def\<{{\langle}}
\def\>{{\rangle}}

\newcommand{\al}{\alpha}

\newcommand{\de}{\delta}
\newcommand{\D}{\Delta}
\newcommand{\e}{\epsilon}

\newcommand{\Lm}{\Lambda}
\newcommand{\w}{\omega}

\newcommand{\s}{\sigma}

\newcommand{\p}{\partial}

\newcommand{\mbh}{\mathbf{h}}

\newcommand{\mbS}{\mathbf{S}}
\newcommand{\mbs}{\mathbf{s}}

\newcommand{\mbz}{\mathbf{z}}
\newcommand{\mbx}{\mathbf{x}}
\newcommand{\bsphi}[1]{\boldsymbol{\phi}}

\newcommand{\hatbf}[1]{\hat{\bf{#1}}}
\newcommand{\bfhat}[1]{\hat{\bf{#1}}}

\newcommand{\mcal}[1]{\mathcal{#1}}

\newcommand{\til}[1]{\tilde{#1}}

\newcommand{\pfrac}[2]{\left(\frac{#1}{#2}\right)}

\newcommand{\bs}[1]{\boldsymbol{#1}}

\newcommand{\Tr}{{\rm Tr}}
\newcommand{\nn}{\nonumber\\}
\newcommand{\up}{\uparrow}
\newcommand{\down}{\downarrow}

\newcommand{\ben}{\begin{equation}}
\newcommand{\een}{\end{equation}}

\newcommand{\onehalf}{\frac{1}{2}}

\newcommand{\ket}[1]{\ensuremath{|{#1}\rangle}}

\begin{document}
\title{High Fidelity Singlet-Triplet ${S}$-${T_-}$ Qubits in Inhomogeneous Magnetic Fields}
\author{Clement H. Wong}
\author{M. A. Eriksson}
\author{S. N. Coppersmith}
\author{Mark Friesen}
\affiliation{Department of Physics, University of Wisconsin-Madison, Madison, Wisconsin 53706, USA}

\begin{abstract}
We propose an optimized set of quantum gates for a singlet-triplet qubit in a double quantum dot with two electrons utilizing the $S$-$T_-$ subspace.  
Qubit rotations are driven by the applied magnetic field    and a field gradient provided by a micromagnet.  
We optimize the fidelity of this qubit as a function of the magnetic fields, taking advantage of ``sweet spots" where the rotation frequencies are independent of the energy level detuning, providing protection against charge noise.   
We simulate gate operations and qubit rotations in the presence of quasistatic noise from charge and nuclear spins as well as leakage to nonqubit states.  
Our results show that, for silicon quantum dots, gate fidelities greater than $99\%$ should be realizable, for rotations about two nearly orthogonal axes. 
\end{abstract}
\date{\today}
\maketitle

\section{introduction}
Electron spins in semiconductor quantum dots are promising qubits because of the long coherence times found in such devices and their potential for scalability~\cite{Zwanenburg:2013p961}.  Single-electron spins have been manipulated by applied AC magnetic fields in both III-V and group-IV devices~\cite{Koppens:2006p766,Pla:2012p541,Veldhorst:2014p981}.  By incorporating micromagnets \cite{PioroLadriere:2008p776} near the quantum dot, AC electric fields can be used for coherent manipulation of single spins~\cite{Nowack:2007p1430,Kawakami:2014p666}.  Magnetic field differences can also be generated by pumping the nuclear spin bath~\cite{Reilly:2008p817,Foletti:2009p903}, and effective fields can be created by electric-field motion in high spin-orbit materials~\cite{Petersson:2012p380}.

By working with two electrons in a double quantum dot, qubits can also be formed from the singlet ($S$) and triplet ($T$) states~\cite{Petta:2005p2180,Foletti:2009p903,Bluhm:2011p109,Maune:2012p344,Wu:2014p11938}.
A magnetic field difference between the quantum dots enables full control of the $S$-$T_0$ subspace by controlling the detuning energy $\epsilon$ between the dots, with the eigenstates varying from $\{\ket{\down\up}, \ket{\up\down}\}$ to \{$\ket{S}$, $\ket{T_0}$\} in different working regimes.  Recently, an alternative two-electron qubit has been studied, consisting of the singlet $\ket{S}$ and polarized triplet $\ket{T_+}$ states for GaAs~\cite{Coish:2005p125337,coish:2007p161302,Ribeiro:2009p216802,Petta:2010p669,Ribeiro:2010p115445,Gaudreau:2012p54,Ribeiro:2013p086804} or the $\ket{S}$ and $\ket{T_-}$ states for Si~\cite{Wu:2014p11938}.  Coherent oscillations have been observed in experiments~\cite{Petta:2010p669,studenikinPRL12}, and theory predicts that such oscillations can be high speed~\cite{Ribeiro:2010p115445,chesi:2014p235311}.
However, the previous work does not resolve whether this qubit can achieve  fidelities high enough to meet the threshold for quantum error correction.

In this work, we determine the optimal working points for pulsed-gating manipulation of the $S$-$T_-$ (or, equivalently, $S$-$T_+$) qubit in a Si double quantum dot.  
The points occur in a regime of magnetic fields and field gradients that has not been elucidated previously.  
The required field gradients are easily achieved with micromagnets~\cite{Kawakami:2014p666}.
Using realistic assumptions about experimental noise derived from recent experiments, we demonstrate that control fidelities in excess of 99\% can be realized in natural abundance Si.  The calculated fidelities are high enough to achieve fault-tolerant operation using surface code error correction~\cite{fowler:2012p032324}.  Interestingly, only one of the optimal operating points is at a charge-noise sweet spot.  The other optimal point is detuned from the second charge-noise sweet spot, in order to avoid leakage driven by the magnetic field difference between the quantum dots.  Using realistic parameters, we find gate speeds of 43~MHz for $X$ rotations and 124~MHz for $Z$ rotations.  These gate speeds can be increased by simultaneously increasing the applied magnetic fields and the interdot tunnel coupling.

This paper is organized as follows.  
In Sec.~\ref{sec:overview}, we review the experimental setup for the $S$-$T_-$ qubit, particularly the required magnetic field configuration.
In Sec.~\ref{ddh}, we develop a  two-electron double dot Hamiltonian including five different spin and charge states.  
In Sec.~\ref{reduced}, we obtain a reduced two-dimensional (2D) Hamiltonian, which spans the qubit subspace.
In Sec.~\ref{qubit}, we describe one and two-qubit gate operations for the $S$-$T_-$ qubit, and we discuss the range of device parameters that should yield high gate fidelities.
In Sec.~\ref{gf}, we describe our simulation techniques and the optimized results for both $X$ and $Z$ rotations, in the presence of both environmental noise and leakage.
In Sec.~\ref{sec:conclusions} we conclude our general discussion.
In Appendix~\ref{tc}, we describe the two tunnel coupling models used in our simulations.
In Appendix~\ref{heff}, we provide details about the derivation of the effective 2D Hamiltonian used in several analytical calculations.
In Appendix~\ref{leakApp}, we describe the analytical solutions for leakage at early times in the dynamical evolution.
In Appendix~\ref{chi}, we provide details about our process fidelity calculations.
In Appendix~\ref{dephase}, we describe our analytical calculations of various dephasing rates, by averaging over quasistatic fluctuating variables, including magnetic and charge noise, and we compare these results with dephasing estimates for a $S$-$T_0$ qubit.
In Appendix~\ref{qEOM1}, we explain why the optimal working point for $Z$ rotations does not occur in the far-detuned limit.

\begin{figure*}[t]
\begin{center}
\includegraphics[width=6.5in]{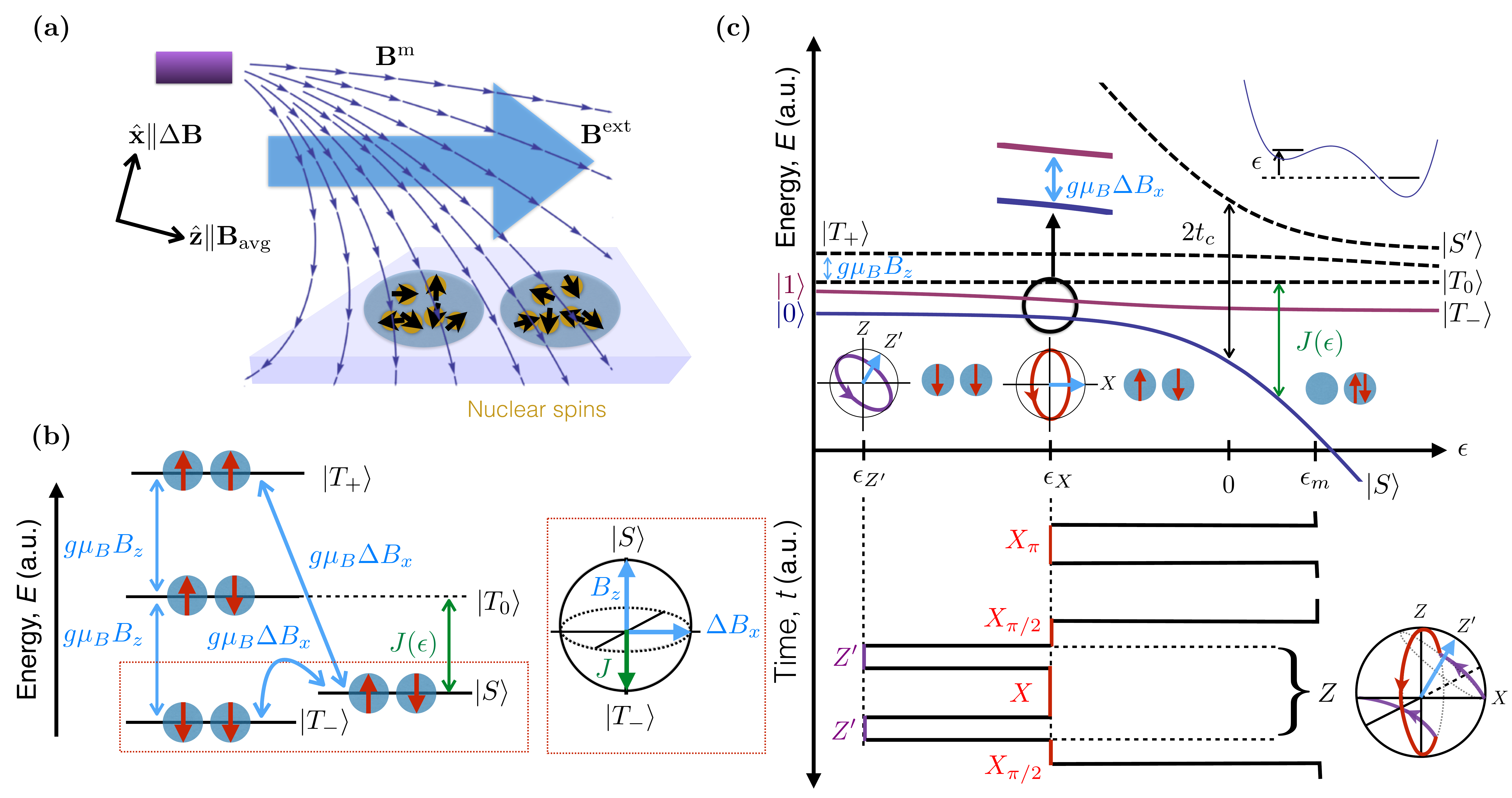}
\caption{(Color online)
(a) Illustration of a nonuniform magnetic field $\mbB^{\rm m}$ provided by a micromagnet (dark purple rectangle) fabricated above a double quantum dot, and a uniform external field $\mbB^{\rm ext}$ (blue arrow).
Random, quasistatic Overhauser fields are also present, due to nuclear spins. 
(b) Singlet-triplet energy diagram, showing the dominant couplings between levels (arrows).
A Bloch sphere representation of the $S$-$T_-$ qubit indicates the rotation axes associated with the different coupling terms. 
(c)  Top:  singlet-triplet energy diagram as a function of detuning $\e$.  
$X$ rotations are performed at a detuning sweet spot (black circle at $\e_X$) where the qubit energy levels are parallel and the splitting is set by $\D B_x$.  
$Z'$ rotations occur in the far-detuned regime ($\e_{Z'}\ll 0$), with a rotation axis $Z'$ tilted slightly away from $Z$ on the Bloch sphere.
Bottom:  illustration of typical pulse sequences for implementing $X$ and $Z$-rotations. 
Measurement of the singlet probability is done at the detuning value $\e_m>0$ in the $(0,2)$ charge state. 
The $Z$ protocol shows a Ramsey pulse sequence where the $Z$ rotation is implemented using a three-step sequence~\cite{Hanson:2007p050502} to correct for the tilt of the $Z'$ axis, as illustrated on the Bloch sphere.}
\label{STQbit}
\end{center}
\end{figure*}
 
\section{$S$-$T_-$ Qubit}
In this section, we first provide an overview of the $S$-$T_-$ qubit.
We then give a detailed description of the Hamiltonian for the 5-level singlet-triplet basis set, as well as an effective 2-level Hamiltonian for the qubit subspace.

\subsection{Experimental overview\label{sec:overview}}

We consider a double dot geometry with a magnetic field gradient generated by a micromagnet, as shown in Fig.~1(a).
There are three contributions to the total field:  the uniform external field $\mbB^\text{ext}$, whose magnitude and direction are assumed to be tunable, the field from the micromagnet $\mbB^{\rm m}$~ \cite{PioroLadriere:2008p776,Wu:2014p11938}, and the slowly varying Overhauser fields $\mbh$, arising from the nuclear spins~\cite{Assali:2011p165301}.
The static fields, $\mbB^\text{ext}$ and $\mbB^{\rm m}$, induce different local fields on the left and right dots, labeled $\mbB_L$ and $\mbB_R$, respectively.
The average field is defined as $\mbB_\text{avg}=(\mbB_L+\mbB_R)/2$, while the the field difference is defined as $\Delta\mbB=\mbB_L-\mbB_R$.
We define the spin quantization axis $\hat{\mbz}$ such that $\mbB_\text{avg}=B_z\hat{\mbz}$.
{The Hamiltonian described in Sec.~\ref{ddh} also includes the local Overhauser fields $\mbh_L$, $\mbh_R$, with $\mbh=(\mbh_L+\mbh_R)/2$ and $\Delta\mbh=\mbh_L-\mbh_R$.}

The main control parameter for the Hamiltonian is the energy detuning $\epsilon$ between the $(0,2)$ and $(1,1)$ charge states, as sketched in the upper inset of Fig.~1(c).
Here, $\epsilon=0$ corresponds to the charging transition.
The relevant energy levels of the two-electron double dot, including energy splittings due to $\Delta\mbB$, are plotted as a function of the detuning in Fig.~1(c).
The energy eigenstates $\ket{0}$ and $\ket{1}$ correspond to the two lowest energy levels in the figure, which are adiabatically (with respect to $t_c$) connected to the qubit logical states $\ket{S}$ and $\ket{T_-}$, the energy eigenstates at the large positive detuning $\e_m$, where the coupling between $\ket{S}$ and $\ket{T_-}$ due to $\D\mbB$  vanishes.
For  $\e<0$, the $\ket{S}$ and $\ket{T_-}$ states hybridize, and diabatic, pulsed gates can drive rotations between them.

Three main mechanisms work to reduce the fidelity of $S$-$T_-$ gate operations:  charge noise, magnetic noise, and leakage to non-qubit sectors of the Hilbert space.
Leakage occurs most readily to states that are closest in energy to the qubit states.
The greatest threat, according to Fig.~1(c), is therefore $\ket{T_0}$.
As is well known from the study of $S$-$T_0$ qubits~\cite{Petta:2005p2180}, this state can become occupied in the presence of a field gradient $\Delta {\bf B}\|\hat{\bf z}$.
{This leakage is minimized when $\Delta\mbB$ and $\mbB_\text{avg}$ are perpendicular.
(For convenience, we define $\Delta\mbB\| \hat{\mbx}$.)}
We assume this alignment, so that the main $S$-$T_-$ leakage channel is avoided, except for unavoidable longitudinal fluctuations of the Overhauser field, $\Delta h_z$.

It has long been known that charge noise can be suppressed in superconducting~\cite{Vion:2002p886} or quantum dot~\cite{Friesen:2002p4619} qubits by operating at a ``sweet spot," where the energy splitting between the qubit states, $E_{01}$, is insensitive to small fluctuations of a control parameter.
For fluctuations of the detuning parameter caused by charge noise, sweet spots are values of $\epsilon$ for which $\partial E_{01}/\partial \epsilon =0$.
The $S$-$T_-$ qubit has two charge-noise sweet spots:  the first occurs at the detuning value $\epsilon_X$, as indicated in Fig.~1(c); the second occurs in the limit $\epsilon\rightarrow -\infty$.
Both locations are candidates for performing gate operations.

We can represent quantum gate operations on a Bloch sphere with $\ket{S}$ at the north pole and $\ket{T_-}$ at the south pole, as indicated in the inset of Fig.~1(b).
In Sec.~\ref{ddh}, we show that $B_z$ generates rotations around the $\hat{\bf z}$ axis, while $\Delta B_x$ generates rotations around the $\hat{\bf x}$ axis.
A third rotation axis $-\hat{\mbz}$ is provided by the exchange coupling $J$, defined as the energy splitting between the $\ket{S}$ and $\ket{T_0}$ states, which is also indicated in the inset of Fig.~1(b).
While the magnetic fields $B_z$ and $\Delta B_x$ remain constant throughout an experiment, the exchange coupling can be tuned electrostatically, through the detuning parameter.
$X$ rotations are achieved by tuning $\epsilon$ to the sweet spot $\epsilon_X$, where $J=g\mu_BB_z$, causing the $Z$ component of the rotation to vanish. 
Here, $g\simeq 2$ is the Land\'{e} $g$ factor in silicon and $\mu_B$ is the Bohr magneton.
We show below that the second sweet spot at $\epsilon\rightarrow -\infty$ is not an optimal working point, and that higher fidelity operations can be achieved at the finite detuning value $\epsilon_{Z'}$.
This point is not perfectly aligned with $\hat{\mbz}$, since $\Delta B_x$ cannot be turned off.
However, short pulse sequences can be used to correct for this misalignment~\cite{Hanson:2007p050502,Shim:2013unpub}.
The pulse sequences considered in this work are shown in the lower portion of Fig.~1(c).

{The detuning sweet spot at $\epsilon_X$ may be enhanced by arranging for $\partial E_{01}/\partial \epsilon \simeq 0$ over as broad a detuning range as possible.
We study this problem in Sec.~\ref{gf} by maximizing the $X$-rotation fidelity in our simulations, finding that the optimal sweet spot occurs at specific values of $B_z$ and $\Delta B_x$.
In practice, $\Delta B_x$ is the most difficult parameter to control experimentally, since it typically depends on the placement of a micromagnet.
Once the desired $\Delta B_x$ has been engineered, the direction of $\Delta \bf B$ determines the $\hat{\bf x}$ axis.
The magnitude and the direction of the external field ${\bf B}^\text{ext}$ must then be chosen to attain the optimal value of $B_z$, while satisfying the requirement $\Delta\mbB\perp\mbB_\text{avg}$.
Practically, such directional control probably requires some trial and error to achieve high accuracy.
However, the process can be facilitated by using a vector magnet.
Indeed, our simulations indicate that optimal operating fields are in the range of 1-10~mT, which could even be achieved via current-carrying wires.
For a given device, the orientation of ${\bf B}^\text{ext}$ only needs to be performed once.
In Appendix~\ref{leakApp}, we show that the proposed $S$-$T_-$ qubit can tolerate misalignments of the field orientation as large as $\D B_z/\D B_x$=10\%, at the optimal working point.}

\subsection{Full double-dot Hamiltonian\label{ddh}}

Here, we describe the full 5D Hamiltonian for two-electron states in a double quantum dot, which yields the energy levels shown in Fig.~1(c).
We begin with a Hubbard Hamiltonian~\cite{Coish:2005p125337,Ribeiro:2009p216802},
\begin{align*}
H&={t_c\over\sqrt{2}}({\vec{c}_{L}}^\dag\vec{c}_{R}+\vec{c}^\dag_{R}\vec{c}_{L})+H_Z(\mbB_i,\mbh_i)\nn
&-\sum_{i=L,R} \mu_i (n_{i\up}+n_{i\down}) +Un_{i\up} n_{i\down}
\end{align*}
where $\vec{c}_i$ is the two-component spinor annihilation operator for electrons on dot $i$=$L$ or $R$, $n_{i\s}=c^\dag_{i\s} c_{i\s}$ is the electron number operator for spin $\s$=$\up$ or $\down$, as defined along the spin quantization axis, $\mu_i$ are electrochemical potentials for the dots, and $U$ is the intradot Coulomb interaction energy \footnote{An interdot Coulomb coupling can be included and absorbed into a redefinition of $\e$.}. The detuning is defined by $\e$=$\mu_L-\mu_R-U$, so that $\e$=$0$ corresponds to the $(1,1)$$\to$$(0,2)$ charge transition. 
The Zeeman Hamiltonian $H_Z$ is defined as $H_Z(\mbB_i,\mbh_i)=g\mu_B\sum_{i=L,R}(\mbB_i+\mbh_i)\cdot\mbS_i$\,,
where $\mbS_i=\vec{c}^\dag_{i}{\bm{\s}}\vec{c}_{i}/2$ are the spin density operators on the left and right dots and $\bm{\s}$ are the Pauli matrices. 

In the parameter regime of our proposed qubit, the tunnel coupling $t_c/\sqrt{2}$ represents the largest energy scale, so it is appropriate to begin our calculation by hybridizing the $\{(1,1), (0,2)\}$ charge basis.
Spin flips due to the Overhauser fields occur only between the $(1,1)$ charge states, and are included explicitly, below.
{Moreover, spin-orbit coupling is very weak in silicon~\cite{Prada:2008p115438}, so mixing of the spin states induced by it is negligible.}
Therefore, charge state hybridization occurs only within the singlet subspace $\{\ket{S(1,1)},\ket{S(0,2)}\}$, whose Hamiltonian is given by
\[H_S=t_c(|S(1,1)\>\<S(0,2)|+{\rm h.c.})-\e|S(0,2)\>\<S(0,2)|\,.\]
Diagonalizing this system yields the hybridized singlet states
\ben
\begin{pmatrix} \ket{S} \\ \ket{S'} \end{pmatrix}
=\begin{pmatrix}\cos\eta &  \sin\eta   \\ -\sin\eta &    \cos\eta\end{pmatrix}
\begin{pmatrix} \ket{S(1,1)} \\ \ket{S(0,2)} \end{pmatrix} ,
\label{mixing}
\een
whose energy eigenvalues are given by
\begin{equation}
\begin{pmatrix} E_S \\ E_{S'} \end{pmatrix}
={t_c} \begin{pmatrix} \tan\eta \\ -\cot\eta \end{pmatrix}
={1\over2}\begin{pmatrix} -\e-\sqrt{4t_c^2+\e^2} \\ -\e+\sqrt{4t_c^2+\e^2} \end{pmatrix}\,. 
\label{eq:eta}
\end{equation}
Here, we have parameterized the admixture of charge states by the mixing angle $\eta$, where $\cos \eta$ and $\sin\eta$ correspond to the amplitudes of the projections of $|S\>$ onto the $(1,1)$ and $(0,2)$ charge states, respectively.
(The magnitudes of $\cos\eta$ and $\sin\eta$ are plotted as a function of detuning in the inset of Fig.~4, in Appendix~A.)
We see that $\ket{S}\rightarrow\ket{S(1,1)}$ when $\e\rightarrow -\infty$ and $\ket{S}\rightarrow\ket{S(0,2)}$ when $\e\rightarrow +\infty$, while $\ket{S'}$ exhibits the opposite asymptotic behaviors.

To evaluate the Hamiltonian in the singlet and triplet basis of spin states defined by
\begin{gather*}
|S\>={|\up\down\>-|\down\up\>\over\sqrt{2}}\,,\quad |T_0\>={|\up\down\>+|\down\up\>\over\sqrt{2}}\,,\\
|T_+\>=|\up\up\>\,,\hspace{0.5in}  |T_-\>=|\down\down\>\,,
\end{gather*}
it is useful to express the Zeeman Hamiltonian in terms of the total spin $\mbS_L+\mbS_R$ and spin difference  $\mbS_L-\mbS_R$ on the two dots.  For the static fields, this yields
\[H_Z(\mbB_i)=g\mu_B\left[{\mbB}_\text{avg}\cdot(\mbS_L+\mbS_R)+{\D{\mbB}\over2}\cdot(\mbS_L-\mbS_R)\right] \,, \]
where the Zeeman Hamiltonian for the Overhauser fields, $H_Z(\mbh_i)$,
is expressed analogously in terms of $\mbh$ and $\Delta {\bf h}$.

Defining the quantization axis $\hat{\bf z}$ such that $\mbB_{\rm avg}=B_z\bfhat{z}$ finally yields a  Hamiltonian, which is projected onto the subspace spanned by the 5D basis set $\{\ket{T_+(1,1)},\ket{T_0(1,1)},\ket{T_-(1,1)},\ket{S},\ket{S'}\}$~\cite{Coish:2005p125337,Taylor:2007p464}:
\begin{widetext}
\ben
H=g\mu_B\left(\begin{array}{cccc|c}
 B_z+h_z & h_+/2 & 0 & \cos\eta \frac{\Delta B_++\Delta h_+}{2 \sqrt{2}}  & -\sin \eta \frac{\Delta B_++\Delta h_+}{2 \sqrt{2}} \\
h_-/2 & 0 & h_+/2& \cos\eta \frac{\Delta B_z+\Delta h_z}{2} & -\sin \eta \frac{\Delta B_z+\Delta h_z}{2}  \\
 0 & h_-/2 & -B_z-h_z & -\cos\eta \frac{\Delta B_-+\Delta h_-}{2 \sqrt{2}} & \sin\eta \frac{\Delta B_-+\Delta h_-}{2 \sqrt{2}} \\
\cos\eta \frac{\Delta B_-+\Delta h_-}{2 \sqrt{2}}  & \cos\eta \frac{\Delta B_z+\Delta h_z}{2} & -\cos\eta \frac{\Delta B_++\Delta h_+}{2 \sqrt{2}} & -J/g\mu_B &0  \\
\hline
-\sin\eta \frac{\Delta B_-+\Delta h_-}{2 \sqrt{2}}  & -\sin\eta \frac{\Delta B_z+\Delta h_z}{2} & \sin\eta \frac{\Delta B_++\Delta h_+}{2 \sqrt{2}} &0& E_{S'}/g\mu_B    \\
\end{array}\right)\,.
\label{H} 
\een
\end{widetext}
Here, we define ${h_\pm=h_x\pm ih_y}$, ${\D B_\pm=\D B_x\pm i\D B_y}$, and ${J=-E_S=(\epsilon/2)}+\sqrt{(\epsilon/2)^2+t_c^2}$, noting that the factors  $\cos \eta$  and $\sin\eta$ associated with the singlet mixing angle $\eta$ appear in the singlet-triplet coupling terms because the triplets only couple to the singlets through the $(1,1)$ charge state. 
Figure~1(b) shows typical singlet-triplet energy splittings near the $S$-$T_-$ crossing, with transitions due to the static field difference $\D\mbB$, as indicated.
The logical $S$-$T_-$ qubit consists of the nearly degenerate subspace of $\ket{S}$ and $\ket{T_-}$ states.
$S$-$T_0$ oscillations, driven by $\Delta B_z$, correspond to leakage outside the qubit subspace.

Finally we note that two prescriptions for the tunnel coupling are employed in the numerical simulations described below:  (i) a constant tunnel coupling, with $t_c$=20~$\mu$eV, and (ii) a detuning-dependent tunnel coupling~\cite{dial:2013p146804} with exponential dependence  $t_c$=$t_0\exp(\epsilon/\epsilon_0)$, and parameters $t_0\sim 20$~$\mu$eV and $\epsilon_0\sim 1$~meV.
The simulation results reported in the main text correspond to case (i), while results for case (ii) are reported in Appendix~\ref{tc} and Table~\ref{texp}; similar results are obtained in both cases.  
{Tunnel couplings of order $t_0=20\,\mu\text{eV}=5$~GHz have been observed in several recent quantum dot experiments~\cite{Simmons:2009p3234,dial:2013p146804}, while values as large as $t_c=60~\mu$eV have also been reported~\cite{studenikinPRL12}}.

\subsection{Reduced Hamiltonian\label{reduced}}

The simulation results reported in this work use the full 5D Hamiltonian of Eq.~(\ref{H}).
However, it is instructive to also consider Hamiltonians of reduced dimension, since they provide intuition and allow us to make analytical progress in some cases.
First, the 5D Hamiltonian can be effectively reduced to 4D by noting that the qubit we propose operates deep in the $(1,1)$ charge regime, where $J/t_c$$\ll$$1$ and $(\cos\eta,\sin\eta)$$\simeq$$(1,-J/t_c)$.
In this limit, the mixing term $\sin \eta$ is very small, and the corresponding probability of leaking into the $\ket{S'}$ singlet is proportional to $\sin \eta^2(|\D\mbB|/E_{S'})^2 $$\approx$$J^4\D\mbB^2/t_c^6$, which
is extremely small.
The physically relevant Hamiltonian is therefore effectively reduced to the upper $4\times4$ block of Eq.~\eqref{H}. 
{As a further simplification, we can also neglect the Rabi flopping terms that couple the different triplet states.
For the parameter regime of interest, these terms lead to effects of order $(h_\pm/B_z)^2\simeq 10^{-6}$, which are very small because the triplet states are split energetically by a large magnetic field~\cite{studenikinPRL12}.
On the other hand, the qubit states $\ket{S}$ and $\ket{T_-}$ are nearly degenerate, so terms involving $\Delta h_\pm$ should not be neglected.}

The 4D Hamiltonian can be further reduced to describe just the 2D subspace of the $S$-$T_-$ qubit.
We first set $\Delta B_z= 0$, as proposed in Sec.~\ref{sec:overview}, to remove the predominant coupling between $\ket{S}$ and $\ket{T_0}$.
We also define the $\hat{\mbx}$ axis such that $\Delta\mbB=\Delta B_x\hat{\mbx}$.
The canonical transformation described in Appendix~\ref{heff} then yields the desired effective Hamiltonian, for which the leading order term is given by~\cite{Taylor:2007p464}
\begin{align}&
H^{(ST_-)}=-g\mu_B \label{H2D} \\ \nonumber & \times
\begin{pmatrix}
J(\e)/g\mu_B& {\cos\eta(\e)\over{2 \sqrt{2}}} (\Delta B_x+\Delta h_+) \\
{\cos\eta(\e)\over{2 \sqrt{2}}}(\Delta B_x+\Delta h_-)  &B_z+h_z
\end{pmatrix} .
\end{align}
In the parameter regime of interest, higher order terms in this expansion are negligibly small.  
Equation~\eqref{H2D} therefore encompasses the full qubit dynamics, except for leakage effects that we show are small in the regime of interest.  

\section{DC pulsed-gate operations\label{qubit}}

The Hamiltonian (\ref{H2D}) enables complete electrical control of the logical qubit  by manipulating the energy detuning $\epsilon$ between the $(0,2)$ and $(1,1)$ charge states, as indicated in Fig.~1(c).  $X$ rotations are performed at the detuning value $\e_X$ defined by $J(\epsilon_X)$=$g\mu_BB_z$, where the $S$-$T_-$ energy levels anticross.
Rotations about an axis close to $Z$ (denoted $Z'$) are performed at a large negative detuning value $\e_{Z'}$, where $J\ll g\mu_B B_z$.   Initialization into the ground state $|S\>$ can be performed deep in the $(0,2)$ charge regime ($\e$$\gg$$0$), where the singlet-triplet coupling is very small because $\cos\eta\simeq 0$.

The simulations reported in Sec.~\ref{gf} suggest that it is important to engineer the broadest possible sweet spot, to suppress the effects of detuning noise.  
Here, we discuss the requirements for achieving a broad sweet spot.
The simplest method to flatten the gap, $g\mu_B\D B_x$, is to increase its size.  
However, according to Eq.~(\ref{H}), the leakage to state $\ket{T_+}$ scales as $(\D B_x/B_z)^2$, so $B_z$ should be simultaneously increased.
Based on such arguments, the full set of requirements for a broad sweet spot is given by
\ben
\s_h\ll g\mu_B\D B_x{\ll} g\mu_B B_z\ll t_c\, ,
\label{optimal}
\een
where $\s_h$ is the variance of the Overhauser field fluctuations.  
The first inequality in Eq.~(\ref{optimal}) ensures that $X$ rotations are much faster than dephasing.
The second inequality suppresses the leakage from $\ket{S}$ to $\ket{T_+}$.  
The final inequality ensures a wide sweet spot  by causing the anticrossing, which occurs at $\e_X\simeq-t_c^2/g\mu_B B_z$ for the parameter range of interest, to occur at large enough negative detunings that even the second derivative $\p^2 E_{01}/\p \e^2\propto(\p J/\p\e)^2$ is very small.
We stress that, while the gradient field should be sufficiently large relative to the nuclear field to achieve high fidelity gate operations, the upper bound set by the tunnel coupling limits the optimal magnitude of $\D B_x$, so that, perhaps counterintuitively, too large  a gradient can degrade qubit fidelity.

The hierarchy of requirements suggested by Eq.~(\ref{optimal}) differs from previous $S$-$T_-$ qubit proposals.
For example, Ref.~[\onlinecite{coish:2007p161302}] proposes to use a \emph{single} spin on one dot as the qubit, and requires a large field gradient, ${B_R^z\ll t_c\lesssim B_L^z}$.
Ref.~[\onlinecite{Ribeiro:2010p115445}] proposes to use small tunnel couplings and field gradients, yielding a narrow sweet spot with $\epsilon_X>0$.
Ref.~[\onlinecite{chesi:2014p235311}] proposes using $\Delta B_x\ll B_z$,  so that leakage effects are suppressed; however they focus on the parameter regime $t_c\lesssim B_z$, where the sweet spot is narrow and $\epsilon_X>0$.
The latter regime is most practical for GaAs devices because large fields are required to combat the effects of Overhauser field fluctuations.
Ref.~[\onlinecite{chesi:2014p235311}] goes on to suggest that $t_c\gg g\mu_BB_z$ would yield a better working regime.
In this paper, we clarify these statements in the context of Si devices, where $t_c\gg g\mu_BB_z$ is not impractical.
We quantify the fidelity levels that can be achieved in a $S$-$T_-$ qubit, for realistic device parameters and realistic noise levels, under the constraints imposed by Eq.~\eqref{optimal}.  
In Appendix~\ref{T2compare}, we further contrast our proposal with the $S$-$T_0$ qubit, which is more sensitive to charge noise.
 
While the present work mainly focuses on single qubit gates, we note that capacitive two-qubit gates can be implemented using the same techniques as $S$-$T_0$ qubits~\cite{Taylor:2005p177,Shulman:2012p202},
because the required capacitive coupling depends only on the orbital charge distribution, not the spin state.  
For example, the entangling component of the dipole-dipole coupling between two qubits, labeled $A$ and $B$, is given by~\cite{Taylor:2005p177}
\begin{align}
H_{AB}=J_{AB}|S(0,2)\>\<S(0,2)|_A\otimes|S(0,2)\>\<S(0,2)|_B
\label{H2}
\end{align}
where $J_{AB}\equiv \D E_c(\p J/\p\e_A)(\p J/\p\e_B)$,  $\D E_c$ is the Coulomb energy difference between the states $\ket{S(1,1)}_A\ket{S(1,1)}_B$ and $\ket{S(0,2)}_A\ket{S(0,2)}_B$, and  $\p J/\p\e=\sin^2\eta$ is proportional to the dipole moment of a given qubit, assuming a constant tunnel coupling.  
The interaction can be turned on by pulsing both qubits to large positive detuning values, where $\sin^2\eta \simeq 1$.  
The result is a CPHASE gate, which has recently been demonstrated in a $S$-$T_0$ qubit system~\cite{Shulman:2012p202}.

\section{Gate fidelities\label{gf}}

\begin{figure*}[t]
\includegraphics[width=6.5in]{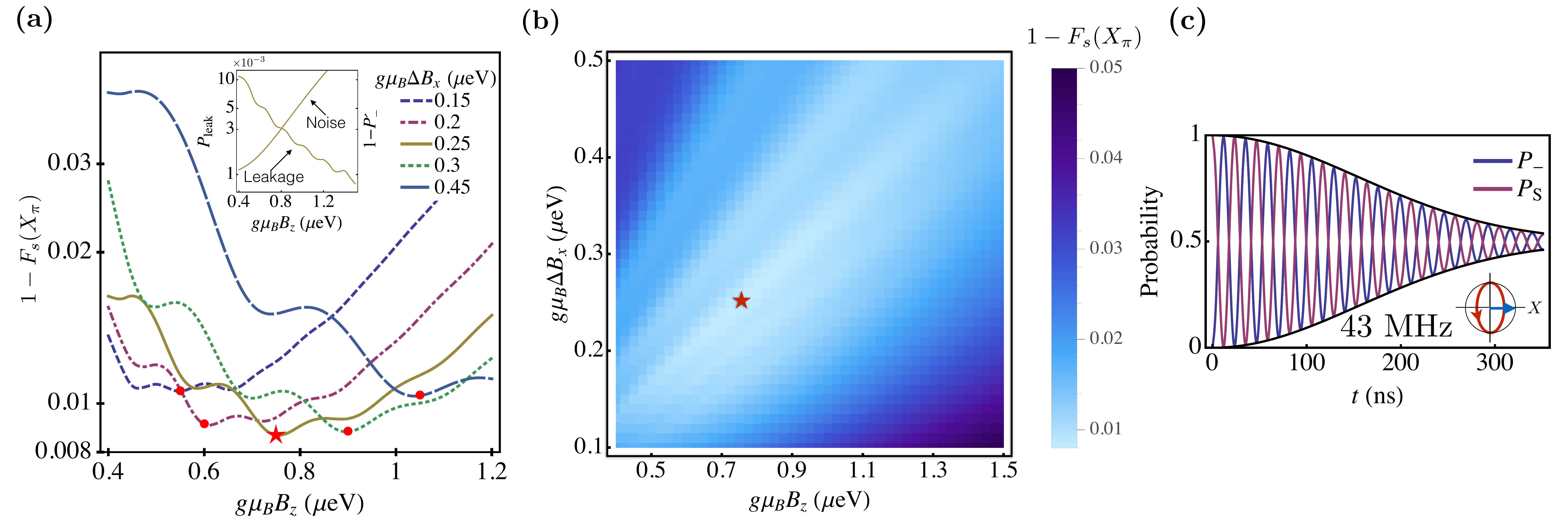}
\caption{(Color online)
Optimization of $X$ rotations.
(a) Semilog plot of  the state infidelity of an $X_\pi$ rotation from $\ket{1}$ to $\ket{0}$, $1-F_s(X_\pi)$, as a function of the applied longitudinal field $B_z$, for several values of the field gradient $\D B_x$, as indicated in the legend. 
Inset:  a similar plot showing the contributions to the state infidelity due to leakage, $P_{\rm leak}$, and the combined effect of detuning and Overhauser field fluctuations, $1-P_-'$, for the case $g\mu_B\Delta B_x$=0.25 $\mu$eV.
(b)  A color density plot of $1-F_s(X_\pi)$ for an $X_\pi$ rotation, as a function of $B_z$ and $\D B_x$.  
The red star indicates the optimal working point $g\mu_B(\D B_x,B_z)$=(0.25,0.75)~$\mu$eV.  
(c)  Larmor oscillations ($X$ rotations), and the corresponding Gaussian decay envelope, $(1\pm e^ {-({t}/{{T_2(X)}})^2})/2$, obtained at the optimal working point, with $T^*_2(X)=\sqrt{2}\hbar/\s_h$.}
 \label{fig:Xrot}
\end{figure*}

\subsection{Simulation procedure}\label{sec:sims}

We now perform  simulations of $S$-$T_-$ qubit rotations, to determine the optimal values for the applied magnetic fields and the tunnel coupling.
 Simulations are performed by numerically integrating the time-dependent Schr\"{o}dinger equation 
\[ i{\hbar}{d c_n\over dt}=\sum_{m=1}^5 H_{nm}(\xi_\al)c_m\,,\]
where $c_n$ are the expansion coefficients of the wave function in the $\{T_+(1,1),T_0(1,1),T_-(1,1),S,S'\}$ basis, $H_{nm}$ are the matrix elements given in Eq.~\eqref{H}, and $\xi_\al$ are noise variables.
 
The main sources of noise in a double dot are detuning fluctuations~\cite{dial:2013p146804,Buizert:2008p1511} and low-frequency Overhauser field fluctuations~\cite{Taylor:2007p464,merkulov:2002p205309},  {which can both be treated quasistatically on the time scales of the qubit dynamics~\cite{Taylor:2007p464}.}
For our simulations, we assume Gaussian distributions over five noise variables, $\xi_\al$$=$$(\de\e,h_z,\D h_{x,y,z})$, where $\de\e$ represents the fluctuation of the detuning from one of its gating positions, $\e_X$ or $\e_{Z'}$, and the remaining variables represent Overhauser field fluctuations away from their average values of zero \footnote{As a check on our approximation, we have compared simulations with and without the Overhauser field fluctuations $h_\pm$, and found that the difference at the optimal working point is $10^{-5}$, consistent with our estimate in Sec.~\ref{reduced}.}.  
The simulations are repeated for nine equally spaced values of a given fluctuation variable in the range 
$\xi_\alpha=-4\sigma_\alpha,\dots, 4\sigma_\alpha$.
The Gaussian average for each component of the density matrix is given by
\begin{equation}
\rho_{nm}=\prod_\al\sum_{k_\al=-4}^4 \sigma_\alpha
[c_n(\xi_\al)c_m^*(\xi_\al) p(\xi_\al)]_{\xi_\al=k_\al\s_\al} \,,
\end{equation}
where $p(\xi_\al)$=$\exp(-\xi_\al^2/2\s_\al^2)/\sqrt{2\pi} \s_\al$ is the Gaussian distribution for variable $\xi_\alpha$.
The variances for the $\delta\epsilon$ and $h_z$ distributions are denoted by $\sigma_\epsilon$ and $\sigma_h$ respectively, while the variance for the $\Delta h_{x,y,z}$ variables is given by $\sqrt{2}\sigma_h$,  as appropriate for uncorrelated noise between the left and right dots. 
Here we adopt the values $\sigma_\epsilon$=5~$\mu$eV~\cite{Maune:2012p344,Wu:2014p11938,shi:2013p075416} and $\sigma_h$=3~neV~\cite{Maune:2012p344,Assali:2011p165301,Wu:2014p11938}, as appropriate for natural Si. 

{
Simulations are performed for a range of control parameters to determine their optimal values for $X$ and $Z$ rotations.
For a given set of control parameters, a Gaussian average is performed over the fluctuating variables, as described above.
We then sweep the control parameters to identify their optimal values, using as our figures of merit the state fidelity of an $X$ rotation from $\ket{S}$ to $\ket{T_-}$, and the state fidelity of a $Z$ rotation from $(\ket{S}+\ket{T_-})/\sqrt{2}$ to $(\ket{S}-\ket{T_-})/\sqrt{2}$.
Finally, we obtain the full process fidelities at the optimal working point for each gate, corresponding to an average over all possible initial states.
In practice, this is accomplished by computing the density matrices for four different initial conditions \cite{NielsenBook}.
The process fidelity is defined as $F_p(\mcal{E})=\Tr[\chi(\mcal{E})\chi(\mcal{E}_0)]$, where $\mcal{E}$ is the final, Gaussian-averaged density matrix from our simulations, and $\mcal{E}_0$ denotes the ideal result, which does not include leakage or noise.
Here, $\chi$ is the process matrix defined by~\cite{NielsenBook}  
\begin{align}
\mcal{E}(\hat{\rho})&=\sum_{mn}\hat{E}_m\hat{\rho}\hat{E}_n\chi_{mn} , \label{eq:tomo}
\end{align}
where we adopt the basis set $\hat{E}_m=\{1,\hat{\tau}_x,-i\hat{\tau}_y,\hat{\tau}_z\}$, and $\hat{\tau}_i$ are Pauli matrices. 
An explicit formula for $\chi$, along with numerical results for the $X_\pi$ and $Z'_\pi$ gates, described in the following sections, is presented in Appendix~\ref{chi}, }

\subsection{$X$ rotations}
We first investigate the fidelity of pulsed $X_\pi$ rotations, using the pulse sequence shown in the lower portion of Fig.~1(c).
The qubit is initialized to state $|S(0,2)\>$ at the detuning value $\epsilon$=$\epsilon_m$,
The qubit is then pulsed via ``rapid adiabatic passage" (RAP)~\cite{Petta:2005p2180}, which is fast compared to the $S$-$T_-$ rotation frequency but slow compared to the tunneling frequency $2t_c/h$.
When we use the optimized magnetic fields, the RAP ramp from $(0,2)$ to $(1,1)$ can be performed so that its contribution to the infidelity due to leakage is negligible (\textless$0.1\%$). 
We then follow the simulation procedure described in Sec.~\ref{sec:sims}, and compute the average probabilities $P_S$, $P_{S'}$, $P_\pm$, and $P_0$ of being in the states $|S\>$, $|S'\>$, $|T_\pm\>$, and $|T_0\>$.
For an $X_\pi$ gate, the state fidelity is defined as $F_s(X_\pi)$=$P_-(\tau_X)$, the probability of reaching the desired target state $|T_-\>$, after a gate evolution period of $\tau_X=h/\sqrt{2}g\mu_B\D B_x$.  
The corresponding infidelity is defined as $1-F_s(X_\pi)$.

Figure~\ref{fig:Xrot} shows the results for the state fidelity of $X$ rotations from $\ket{1}$ to $\ket{0}$, as a function of the magnetic fields $B_z$ and $\Delta B_x$, corresponding to case (i), the constant tunnel coupling model.
We identify the optimal working point, marked by a star in Fig.~2(b), as $g\mu_B\Delta B_x$=0.25~$\mu$eV ($\D B_x$=$1.5$ mT) and $g\mu_BB_z$=0.75~$\mu$eV ($B_z$=$4.5$ mT), corresponding to an $X_\pi$ rotation speed of 43~MHz.
{At this optimal point, we calculate a full process fidelity of $F_p(X_\pi)=99.3\%$, which is slightly higher than the optimized state fidelity, and somewhat higher than fidelities observed in recent experiments~\cite{Ribeiro:2013p086804}.}
The long-lived Larmor oscillations shown in Fig.~\ref{fig:Xrot}(c) are also obtained at the same optimal point.

\begin{figure}[t]
\includegraphics[width=2.2in]{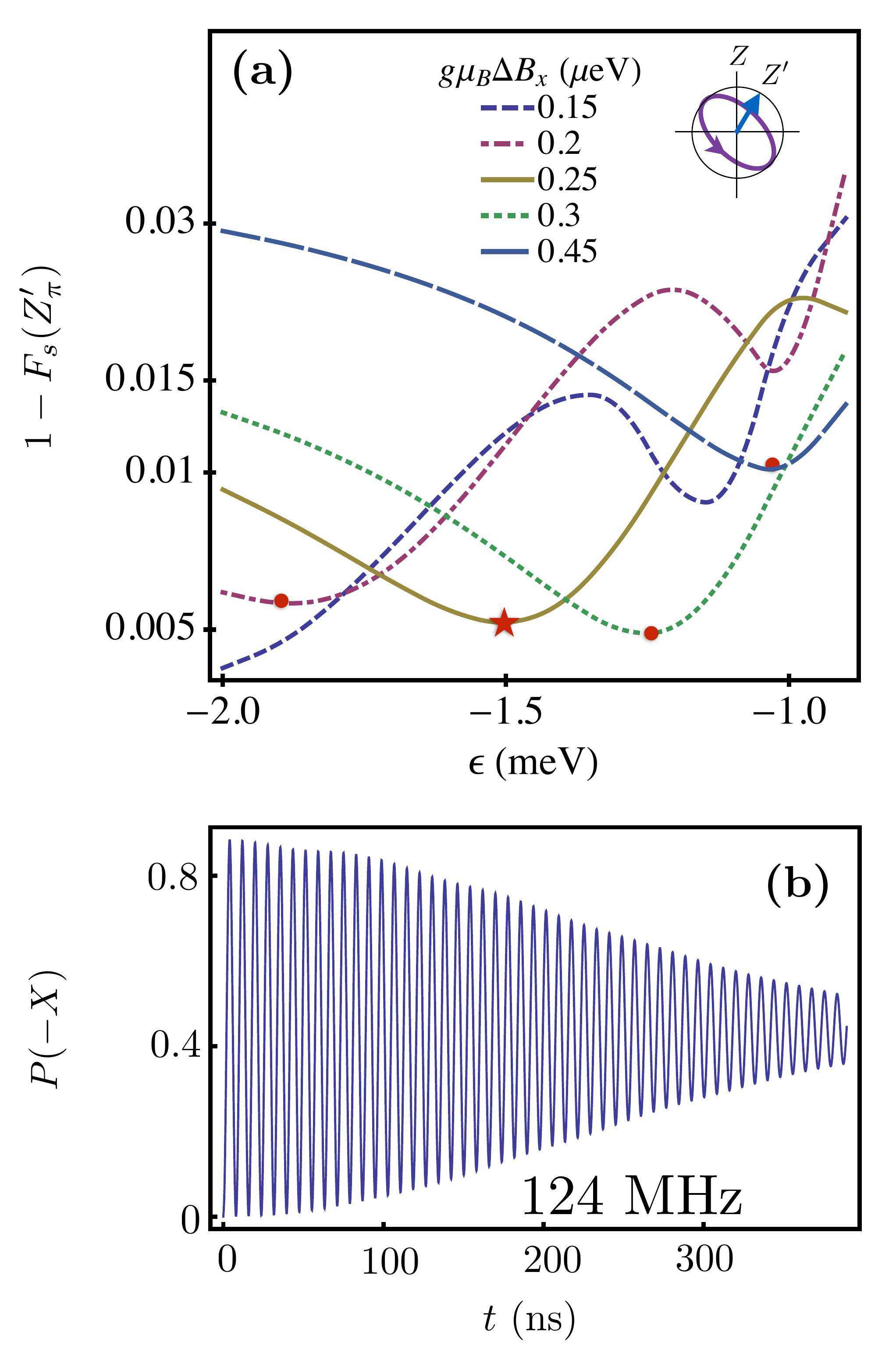}
\caption{(Color online)
Optimization of $Z'$ rotations.
(a) Semilog plot of  the state infidelity of a $Z'_\pi$ rotation, $1-F_s(Z'_\pi)$, as a function of the detuning, for nearly-optimal values of $g\mu_B\D B_x$  and $g\mu_BB_z$.
The red markers indicate the correspondence with curves in Fig.~\ref{fig:Xrot}(a), with the star indicating the optimal working point.
(b)  $Z'$ rotations performed at the starred point in (a), for $\e_{Z'}$=$-1.5$ meV.}
 \label{fig:Zrot}
\end{figure}

Figure~\ref{fig:Xrot}(a) shows that the infidelity goes through a minimum as a function of $B_z$, for a given value of $\Delta B_x$, which results from a competition between leakage and noise. 
To the left of the minimum, the infidelity is dominated by leakage.
To see this, we plot the total leakage probability $P_{\rm leak}$=$P_++P_0+P_{S'}$ in the inset.
Here, the small oscillations are caused by leakage to the state $\ket{T_+}$, which arises from the $\D B_x$ coupling term \footnote{Since the leakage to state $\ket{T_+}$ is coherent, it could potentially be used to form a qutrit in this system; however this would require more control parameters than are available in the present setup.}.
To the right of the minimum, the infidelity is dominated by the exchange noise, ${\de J=(\p J/\p\e)\de\e}$.
To see this, we obtain an approximate analytic solution for the qubit dynamics from the 2D Hamiltonian, Eq.~\eqref{H2D}, which includes detuning fluctuations, but no leakage.
Details of the calculation are given in Appendix~\ref{qEOM}, yielding the final state ($T_-$) probability for an $X_\pi$ rotation 
\ben
P_-'(\tau_X)
=\onehalf\left[{1+e^{-(\tau_X/{ T_2^*(X)})^2}(1-\s^2_\perp/b^2)}\right]\,,
\label{Pma}
\een
where $T_2^*(X)=\sqrt{2}\hbar/\sigma_h$, ${\s_\perp^2={\s_J^2+\s_{h}^2}}$, and  $b=g\mu_B|\D B_x|/\sqrt{2}$. 
The only dependence on $B_z$ in Eq.~(\ref{Pma}) comes from the  exchange noise, whose variance is given by ${\s_J=\s_\e(\p J/\p\e)_{\e=\e_X}}$.
This contribution to the infidelity, $1-P_-'$, which is dominated by exchange noise, is also plotted in the inset of Fig.~2(a).
The sum of the curves in the inset matches the numerical simulations in the main panel and explains the crossover between leakage and noise-dominated behavior.

The competition between leakage and noise suggests a strategy for improving the fidelity of $X_\pi$ rotations:  suppress the charge noise while keeping the leakage constant.
{In Sec.~\ref{qubit}, we noted that leakage into state $\ket{T_+}$ is proportional to $(\D B_x/B_z)^2$, which we want to keep constant.
We can also obtain a scaling relation for the exchange noise ($1-P_-'$) by noting that the main dependence of Eq.~(\ref{Pma}) on $B_z$ arises from the $(\sigma_J/b)^2$ term.
From the definition of $\sigma_J$, and the relation $J\simeq -t_c^2/\epsilon$, which is valid in the vicinity of the working point $\epsilon_X$, defined by $J=g\mu_B\Delta B_x$, we can estimate that $\s_J$$\simeq$$(g\mu_B B_z/t_c)^2\s_\e$$\simeq$7~neV, and the exchange noise contribution to infidelity as
\begin{equation}
(1-P_-')_{\de J}\sim \s_\e^2 B_z^4/t_c^4\D B_x^2\, . \label{eq:scaleP-}
\end{equation}
If $(\D B_x/B_z)^2$ is held fixed, the exchange noise can therefore be suppressed by  reducing $B_z/t_c$, as consistent with Eq.~(\ref{optimal}).}

The optimal value of $g\mu_B\Delta B_x=0.25$~$\mu$eV suggested by our simulations is slightly larger than typical Overhauser fields observed in GaAs dots with random nuclear polarization~\cite{Assali:2011p165301} ($\sim$0.1~$\mu$eV), and much larger than the Overhauser fields in Si dots~\cite{Assali:2011p165301} ($\sim$3~neV).  
Experimentally, values of $\Delta B_x$ as large as 30~mT (3.5~$\mu$eV) have been achieved using micromagnets~\cite{PioroLadriere:2008p776}.
By increasing $\Delta B_x$ to this range, while satisfying the requirements of Eq.~(\ref{optimal}), we can expect to achieve optimal fidelities similar to those in our simulations, with a ten-fold increase in gate speed. 

Finally, we have studied the important role that tunnel coupling plays in determining the optimal $X_\pi$ gate fidelity by repeating our simulations with a smaller tunnel coupling, $t_c$=$10~\mu$eV, instead of the $t_c$=$20~\mu$eV coupling that was used for all the other results described above.
In this case, we obtain a maximum process fidelity of ${F_p(X_\pi)=98.4\%}$ (instead of 99.3\%), corresponding to the optimal working point $g\mu_B(\D B_x,B_z)$=$(0.15,0.3)~\mu$eV.   

\subsection{$Z'$ rotations}
In the laboratory, it is not feasible to tune the magnetic fields differently for $X$ and $Z'$ rotations.
However, once the $\Delta B_x$ and $B_z$ that optimize the $X$ rotations are fixed, we are still free to adjust the value of $\epsilon_{Z'}$ to optimize $Z'$ rotations.
As noted in Sec.~\ref{sec:overview}, the rotation axis $\hat{\bf z}'$ is tilted away from $\hat{\bf z}$ by the angle $\theta$=$\tan^{-1}[\D B_x/\sqrt{2}(B_z-J(\e_{Z'}))]$ in the $x$-$z$ plane.
If desired, a true $Z$ rotation could be implemented via a three-step pulse sequence, provided that $\theta$\textless$45^\circ$~\cite{Hanson:2007p050502}.
In turn, this sequence can be incorporated into longer sequences, like the Ramsey sequence, $X_{\pi/2}$-$Z_\pi$-$X_{\pi/2}$, shown in Fig.~1(c).
Here, we simulate just the $Z'(\pi)$ portion of the sequence.
Beginning with the initial state $|X\>$=$(|S\>+|T_-\>)/\sqrt{2}$, on the equator of the $S$-$T_-$ Bloch sphere, we suddenly pulse the detuning to $\e_{Z'}$ and evolve the system for a $\pi$-rotation period.
We then compute the state fidelity $F_s(Z'_\pi)$=$P(-X)/V$, where $P(-X)$ is the probability of reaching the final state $\ket{-X}$ and   $V=\cos^2\theta$ is the visibility, defined as the maximum amplitude for an ideal $Z'$ rotation \footnote{Here, we only consider $\epsilon_{Z'}<\epsilon_X$, to reduce the effects of charge noise.}.

The results of our simulations of the $Z'$ gate are shown in Fig.~\ref{fig:Zrot}(a).
Here, we have used the same $(\Delta B_x,B_z)$ combinations indicated by markers in Fig.~\ref{fig:Xrot}(a), which yield nearly optimized fidelities for $X_\pi$ rotations, and we perform our optimizations over the detuning parameter $\epsilon_{Z'}$.
For the particular combination $g\mu_B(\D B_x,B_z)$=(0.25,0.75)~$\mu$eV,
we obtain an optimal state fidelity of about $99.5\%$ at $\e_{Z'}$$\simeq$$-1.5$~meV, corresponding to a gate frequency of 124~MHz.  
Other combinations of $\Delta B_x$ and $B_z$ can achieve results with higher $Z'$ fidelities, but lower $X$ fidelities.
{The full process fidelity at the optimal working point marked by a star in Fig.~\ref{fig:Zrot}(a) is computed to be $F(Z'_\pi)=99.9\%$.}
We can compute $T_2^*$ at this same working point using Eq.~(\ref{T2}) in the Appendix, based on quasistatic fluctuations of the detuning and Overhauser fields, obtaining $T_2^*(Z')\simeq300$~ns.
This is consistent with the long-lived $Z'$ oscillations shown in Fig.~\ref{fig:Zrot}(b), while the observed visibility of 89\% is consistent with the rotation tilt angle of $\theta$=$19.5^\circ$ obtained at the optimal working point.

Similar to $X_\pi$ rotations, the optimal value of $\e_{Z'}$ is determined by a competition between detuning noise and leakage, {and occurs exclusively in the regime $-\infty<\e_{Z'}<\e_X$.}
Denoting the optimal gate location by $\e_{Z'}^*$,  {we find that detuning noise quickly suppresses the fidelity when ${\e_{Z'}>\e_{Z'}^*}$}, while leakage dominates when  ${\e_{Z'}<\e_{Z'}^*}$.
Leakage is particularly evident in the asymptotic limit $\epsilon_{Z'}\rightarrow -\infty$, where exact analytic solutions are available, as described in Appendix~\ref{qEOM1}.
Here, the coherent oscillations arising from leakage into the state $\ket{T_+}$ are far more prominent than at the optimal working point $\e_{Z'}=\e_{Z'}^*$. 
 
\section{Conclusions and outlook}\label{sec:conclusions}

In conclusion, we have investigated in detail a singlet-triplet qubit in the $S$-$T_-$ subspace, for which all rotation frequencies are set by the applied magnetic fields.   
By simulating a quasistatic noise model, we have shown that in the regime $\D B_x\lesssim B_z\ll t_c$, the qubit is well protected from detuning noise, due to presence of a broad sweet spot at the $S$-$T_-$ anticrossing.
The magnetic field gradients needed to achieve such a sweet spot are smaller  than those considered in previous proposals~\cite{Ribeiro:2010p115445,chesi:2014p235311}, due in part to using Si as a substrate, so that the effects of Overhauser fields are reduced.
The required fields are relatively easy to produce in the laboratory by means of micromagnets and a tunable external field, yielding fidelities that should exceed $99\%$ for rotations around two nearly orthogonal axes.

The fidelities predicted here depend on the input parameters used in the simulations, and they can potentially be enhanced in several ways.  
First, charge noise can be reduced by special sample fabrication and preparation~\cite{Buizert:2008p1511}.
Further improvements in materials could also reduce the charge noise.
Second, the leakage and dephasing mechanisms considered here can both be suppressed by  increasing the tunnel coupling and then re-optimizing the magnetic fields.
Third, the nuclear noise can be reduced by using isotopically purified $^{28}$Si.
We estimate that the dominant dephasing mechanism would switch from Overhauser to detuning noise at the level of 99.5$\%$ isotopic purification, corresponding to $\sigma_h$\textless$0.2$ neV.  
(See Appendix \ref{nuclear}.)
{Under these conditions, assuming $t_c=60~\mu$eV and the optimal working point $g\mu_B(\D B_x,B_z)=(0.1,0.9)~\mu$eV, our model predicts an $X_\pi$ gate fidelity of $F_p(X_\pi)=99.9\%$.}
For materials like GaAs or InGaAs, where spin-0 isotopic purification is not an option, the fluctuation spectrum can be narrowed by nuclear polarization~\cite{Foletti:2009p903,brataas:2011p045301,Petta:2008p067601,Reilly:2008p817}.
In such materials, $\D B_x$ can also be controlled via nuclear polarization, or by making use of a large spin-orbit coupling~\cite{rudner:2010p041311}.

While the analysis here has focused on DC pulsed gates, AC resonant gates have some advantages~\cite{Shulman:2014p5156}.
In particular, they allow all qubit operations to be performed at the sweet spot $\e_X$.
Recent experiments performed at the sweet spot of a charge qubit show significant improvements in fidelity  for AC gates~\cite{kim:2015p243} compared to DC gates~\cite{Petersson:2010p246804,shi:2013p075416}.
Similar improvements are observed in the quantum dot hybrid qubit~\cite{Koh:2012p250503}, where theory indicates that better fidelity should be obtained for AC gates~\cite{Koh:2013p19695}; this is confirmed in experiments by comparing AC gates~\cite{Kim:2015unpub} and DC gates~\cite{Kim:2014p70}.

This work was supported in part by NSF (PHY-1104660), NSF (DMR-1206915),  ARO (W911NF-12-0607), UW-Madison Bridge Funding (150 486700 4), and by the Intelligence Community Postdoctoral Research Fellowship Program.  
The views and conclusions contained in this document are those of the authors and should not be interpreted as necessarily representing the official policies or endorsements, either expressed or implied, of the U.S. Government.

\appendix

\section{Detuning dependent tunnel coupling\label{tc}}
\begin{figure}[t]
\includegraphics[width=2.5in]{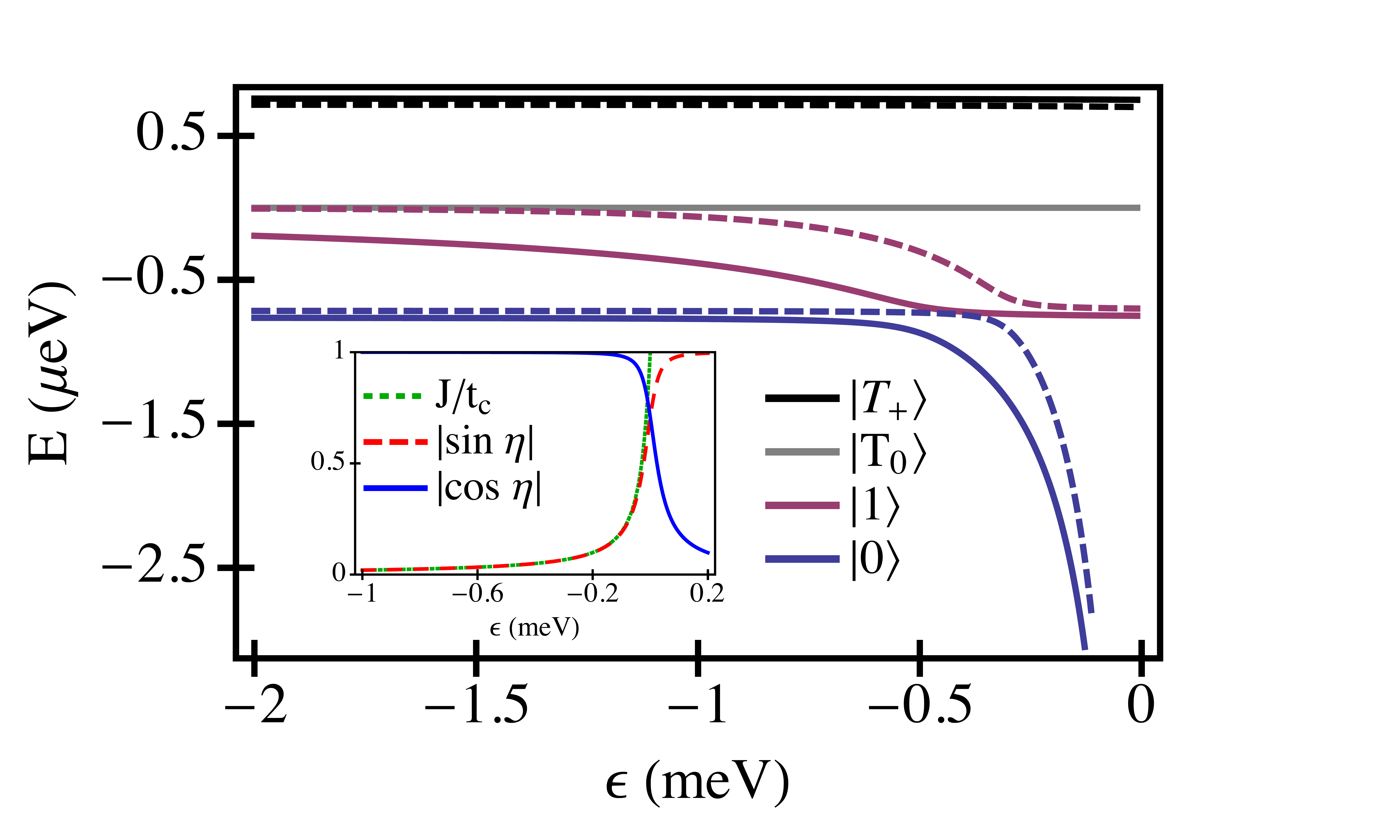}
\caption{(Color online)  
Singlet-triplet energy diagrams, including the qubit states $\ket{0}$ and $\ket{1}$ and leakage states $\ket{T_0}$ and $\ket{T_+}$, as a function of the detuning, for two different tunnel coupling models.  
Here, state $\ket{S'}$ lies outside the range of the plot.
Solid lines:  case (i), the constant tunnel coupling model, with $t_c$=$20~\mu$eV and magnetic field values $g\mu_B(\D B_x,B_z)$=$(0.25,0.75)~\mu$eV, which were optimized as described in the main text.  
On the right-hand side of the anticrossing, the qubit states correspond to $|0\>$$\simeq$$|S\>$ and $|1\>$$\simeq$$|T_-\>$. 
Dashed lines:  case (ii), the detuning-dependent tunnel coupling model, $t_c(\e)$=$t_0\exp(\e/\e_0)$, with $t_0=20$~$\mu$eV and $\epsilon_0=1$~meV.
Here too, the magnetic field parameters $g\mu_B(\D B_x,B_z)$=$(0.3,0.7)~\mu$eV were optimized to achieve a high $X_\pi$ gate fidelity. 
Inset:  the singlet mixing terms $(|\cos\eta|,|\sin\eta|)$, from Eq.~(\ref{mixing}).
Note that $|\sin \eta|\simeq J(\e)/t_c$, the effective exchange energy, to a very good approximation.}
\label{energy}
\end{figure}

In the main text, we considered two models for the tunnel coupling.
Case (i) corresponds to the constant tunnel coupling model, which has been used by many workers~\cite{Petta:2005p2180,Taylor:2005p177}.
The results shown in Fig.~2 use the constant value $t_c=20$~$\mu$eV.
In this Appendix, we consider case (ii), for which the tunnel coupling takes the exponential form $t_c(\e)$=$t_0\exp(\e/\e_0)$.
Recent experiments suggest that such dependence may occur in both GaAs~\cite{dial:2013p146804} and Si systems~\cite{Maune:2012p344,Wu:2014p11938}.
Fitting the exchange energy data from Fig.~5 of Ref.~[\onlinecite{Maune:2012p344}] to the form $J$=$t_c(\e)^2/|\e|$, which is valid in the weak tunneling limit ($t_c$$\ll$$|\e|$), we obtain the estimate $\e_0\simeq10$~meV.
Similarly, a fit to the data in Ref.~[\onlinecite{Wu:2014p11938}] suggests that $\e_0=1$-10~meV.
Here, we consider a range of $\epsilon_0$ values, as indicated in Table~\ref{texp}, to explore their effect on our main results. 
To facilitate a comparison with case (i), we adopt $t_0=20$~$\mu$eV.

Energy level diagrams for cases (i) and (ii) are shown in Fig.~\ref{energy}, with solid and dashed lines, respectively.
In each case, the results were obtained using magnetic fields that optimize the given model.
Note that the highest energy level $\ket{S'}$ lies outside the range of the plot.
For the relatively small value of $\epsilon_0=1$~meV assumed in this figure, the qubit levels for case (ii) quickly approach their asymptotic values, and deviate from case (i).
As a result, the working points $\e_X$ and $\e_{Z'}$ for case (ii) are both shifted to the right, compared to case (i).  
In contrast, for large values of $\epsilon_0$, the two models are nearly identical over the entire parameter range of interest.

\begin{table}[t]
\begin{tabular}{cccc}
\hline\hline
$\e_0~({\rm meV})$&$g\mu_B\D B_x~(\mu{\rm eV})$&$ g\mu_BB_z~(\mu{\rm eV}) $&$F_p(X_\pi)~(\%) $\\
\hline
 $1$&$0.3$ &$ 0.7$ &$ 98.5$ \\
 $10$&$0.25$ & $0.75$ &$ 99.2$ \\
 $100$&$0.25$ & $0.75 $& $99.3$\\
\hline\hline
\end{tabular}
\caption{Process fidelities of the $X_\pi$ gate, $F_p(X_\pi)$, obtained at the indicated optimal magnetic fields, for the detuning-dependent tunnel coupling model $t_c(\e)$=$t_0e^{\e/\e_0}$, with $t_0$=$20~\mu$eV and three different values of $\e_0$.}
\label{texp}
\end{table}

Table~\ref{texp} shows the optimal $X_\pi$ process fidelities $F_p(X_\pi)$ obtained for case (ii), using magnetic fields optimized separately for each value of $\epsilon_0$.
At ${\e_0=1}$ meV, we obtain a fidelity that is slightly lower than for case (i), which was reported in the main text. 
However for ${\e_0=10}$ meV, the optimized fidelity exceeds 99\%,
and for $\e_0$$\simeq$100 meV, we recover the fidelity ${F_p(X_\pi)=99.3\%}$ of case (i).


\section{Effective $S$-$T_-$ Hamiltonian\label{heff}}
The 2D Hamiltonian $H^{(ST-)}$, presented in Eq.~(\ref{H2D}), is obtained by isolating the $S$-$T_-$ subsector of the full 5D Hamiltonian.
In this Appendix, we formally derive the effective 2D Hamiltonian for the $S$-$T_-$ qubit using nearly degenerate perturbation theory~\cite{lowdinJCP51}.
The resulting Hamiltonian $H_\text{eff}$ includes corrections to $H^{(ST-)}$, defined by
\begin{equation}
H_{\rm eff}=H^{(ST_-)}+\de H^{(ST_-)}\, . \label{eq:Heff}
\end{equation}
The correction term $\delta H^{(ST_-)}$ arises at second order in the perturbation~\cite{Coish:2005p125337}, and accounts for virtual transitions into the leakage states $\ket{T_0}$ and $\ket{T_+}$

We begin with the full 5D Hamiltonian given in Eq.~(\ref{H}).
As discussed in the main text, the excited $(0,2)$ charge state $\ket{S'}$ is well split off from the $(1,1)$ charge manifold in the regime of interest, so we only need to consider the upper $4\times4$ block of Eq.~\eqref{H}.  
We now block diagonalize the $S$-$T_-$ subspace, yielding a correction term given by~\cite{lowdinJCP51}
\ben
\de H^{(ST_-)}=H_{PQ}\frac{1}{E-H_{QQ}}H_{QP}\,,
\label{Heff}
\een
where $H_{PP}$=$PHP$=$H^{(ST_-)}$, $H_{QQ}$=$QHQ$,  $H_{QP}$=$QHP$, and $H_{PQ}$=$PHQ$.
Here, $P$=$\sum_i|p_i\>\<p_i|$ is the projection operator onto the  $S$-$T_-$ subspace with state labels $p_i$,  and $Q$=$\sum_i|q_i\>\<q_i|$ is the projection operator onto  the $T_0$-$T_+$  subspace 
with state labels $q_i$.  To leading order in the correction,  $E$ corresponds to the average energy eigenvalue of $H^{(ST_-)}$.  At the $S$-$T_-$ anticrossing, we have $E$=$E_S$=$E_{T_-}$=$-B_z$. Thus, neglecting corrections of order $h_\pm/B_z$, the energy denominator is given by 
\begin{widetext}
\begin{align*}
\frac{1}{E-H_{QQ}}&=\begin{pmatrix} 1/(E_S-E_{T+}) & 0 \\ 0 & 1/(E_S-E_{T_0}) \end{pmatrix}
=\begin{pmatrix} -1/(2g\mu_BB_z) & 0 \\ 0 & -1/(2g\mu_BB_z) \end{pmatrix} \, .
\end{align*}
From Eq.~\eqref{H}, we have 
\begin{align*}
H_{PQ}&=\begin{pmatrix}
\<S|H|T_+\> & \<S|H|T_0\> \\
 \<T_-|H|T_+\>&  \<T_-|H|T_0\>
\end{pmatrix}
=\begin{pmatrix}
\cos \eta  \frac{\Delta B_+   }{2 \sqrt{2}} & \cos\eta \frac{\Delta B_z}{2} \\
 0 & \frac{h_+}{2} 
\end{pmatrix}\,,
\end{align*}
and  $H_{PQ}$=$H_{QP}^\dag$, where $\eta$ was defined in Eq.~(\ref{eq:eta}).  
We then find that
\[\de H_{ST}=-\frac{g\mu_B}{4 B_z}\begin{pmatrix}
 \cos^2\eta \left[(\Delta B_++\Delta h_+) (\Delta B_-+\Delta h_-)/4+(\Delta B_z+\Delta h_z)^2\right] & \quad \cos\eta\, h_- (\Delta B_z+\Delta h_z) \\
 \cos\eta\,  h_+ (\Delta B_z+\Delta h_z) & \quad h_- h_+ \\
\end{pmatrix}\,.\]
\end{widetext}
Since the micromagnet field gradients, $\Delta \bf B$, are much larger than the Overhauser fields, $\bf h$ or $\Delta \bf h$, the leading order correction to $H_\text{eff}$ is given by
\ben
\delta H_{ST}^{(0)}=-g\mu_B\cos^2\eta\frac{(\Delta B_+ \Delta B_-/4+\Delta B_z^2)}{4B_z} {(1+\hat{\tau} _z)\over2}\,.
\label{dH}
\een
This term slightly shifts the location where $X$ rotations are performed, $\epsilon_X$,
which is defined by Tr$[\hat{\tau}_z H_{\rm eff}(\e_X)]$=$0$.  
However, in the optimal operating regime of the $S$-$T_-$ qubit, we find that this correction is negligible.

\section{Leakage\label{leakApp}}

\begin{figure}[t]
\begin{center}
\includegraphics[width=2.7in]{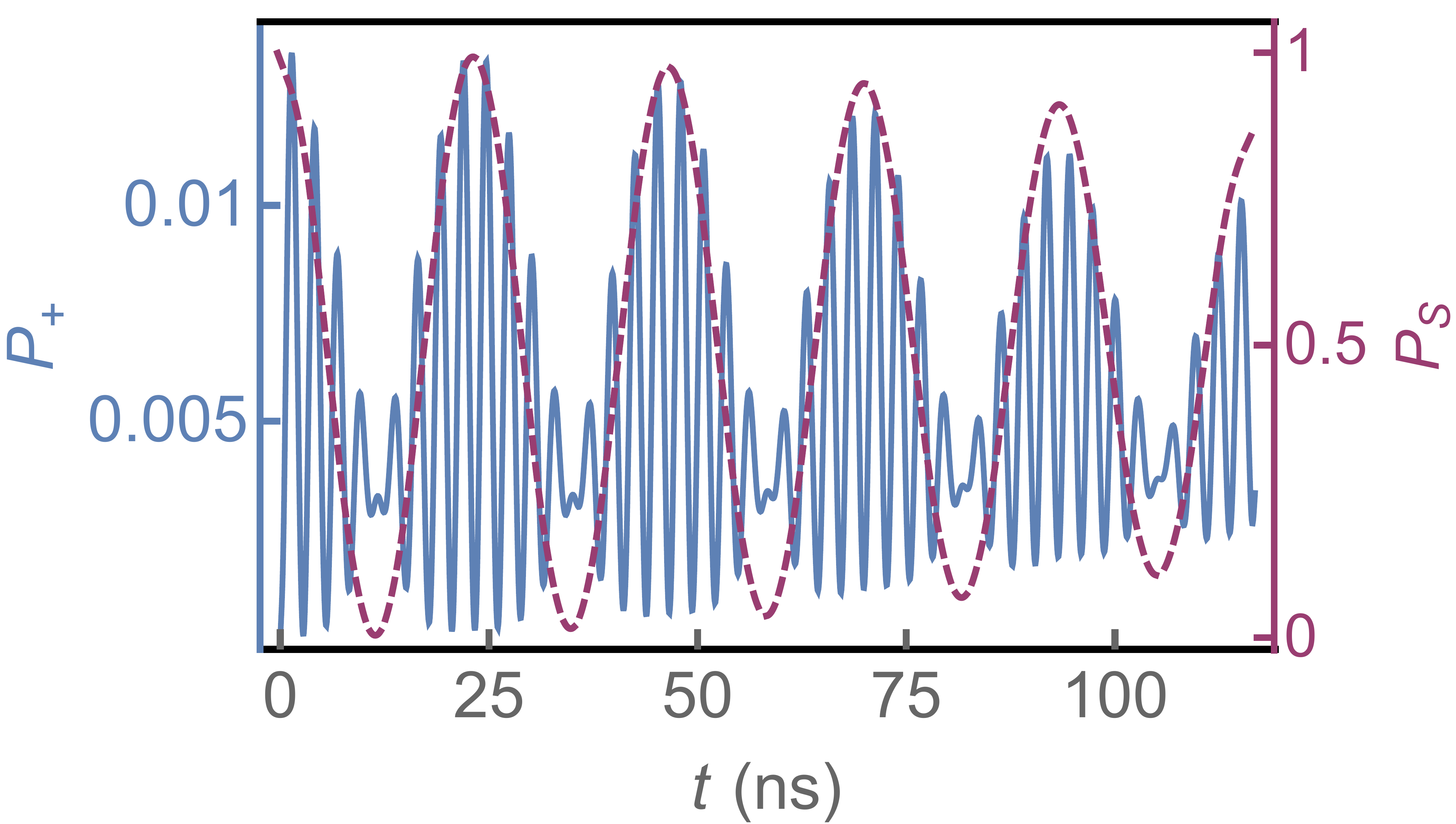}
\caption{(Color online)
Probability $P_+$ of leaking into state $|T_+\>$  as a function of time $t$ during $X$ rotations, for the optimal field values ${g\mu_B(\D B_x,B_z)=(0.25,0.75)~\mu}$eV.  The envelope of the oscillations closely follows the probability $P_S$ of occupying state $|S\>$, as indicated by the maroon, dashed line (right-hand axis).}
\label{leak}
\end{center}
\end{figure}


In this Appendix, we investigate transitions into the leakage states $\ket{T_+}$, $\ket{T_0}$, and $\ket{S'}$.  Our numerical simulations indicate that typical probabilities for occupying the states $\ket{T_0}$ and $\ket{S'}$ are of the order  $P_0$$\simeq$$10^{-5}$ and $P_{S'}$$\simeq$$10^{-10}$, respectively, which can be safely neglectd.
$\ket{T_+}$ is therefore the predominant leakage state, since it is driven by the same process as the desired rotations between $\ket{S}$ and $\ket{T_-}$.

The probability of occupying $\ket{T_+}$ during $X$ rotations is plotted in Fig.~\ref{leak} for the magnetic field combination $g\mu_B(\D B_x,B_z)$=$(0.25,0.75)~\mu$eV, with the qubit initialized to $\ket{S}$.
The leakage exhibits a beating pattern:  the fast oscillations occur at the leakage frequency, $2g\mu_BB_z/h$, while the low-frequency envelope is commensurate with the $S$-$T_-$ rotations, at the qubit frequency $g\mu_B\D B_x/\sqrt{2}h$.   
The latter modulation is specifically proportional to the singlet probability because $\ket{T_-}$ does not couple directly to $\ket{T_+}$.  
We verify this by plotting the singlet probability $P_S$ in Fig.~\ref{leak} with a dashed line that clearly follows the main features of the $\ket{T_+}$ envelope. 
For short times, the full probability is well described by the Rabi formula for off-resonant transitions from $\ket{S}$ to $\ket{T_+}$~\cite{sakuraiQM}:
\ben
P_+\propto P_S\pfrac{\D B_{x}/\sqrt{2}}{2B_z}^2\sin^2(g\mu_B B_z t/\hbar)\,.
\label{Pleak}
\een
The factor $P_S$ takes into account the envelope of leakage oscillations, which go through a point of minimum amplitude at the end of each $X_\pi$ rotation, when the $\ket{S}$ state is not occupied.

Fig.~\ref{leak} shows  a leakage probability of ${P_+\simeq0.003}$ for the $X_\pi$ gate.  This is actually a local maximum of the minima in the leakage oscillations.  
For another gate, such as  $X_{\pi/2}$, Fig.~\ref{leak} suggests that even lower leakage probability can be found. 
Such an optimal point occurs for magnetic fields values such that the rotation and leakage frequencies are commensurate, so that the leakage probability is at a minimum at the end of the gate period.  
Our optimization procedure already finds such optimal points,  corresponding to the local minima in infidelity shown in Fig.~2 (a) and (b).  

\begin{figure*}[t]
\begin{center}
\includegraphics[width=4.5in]{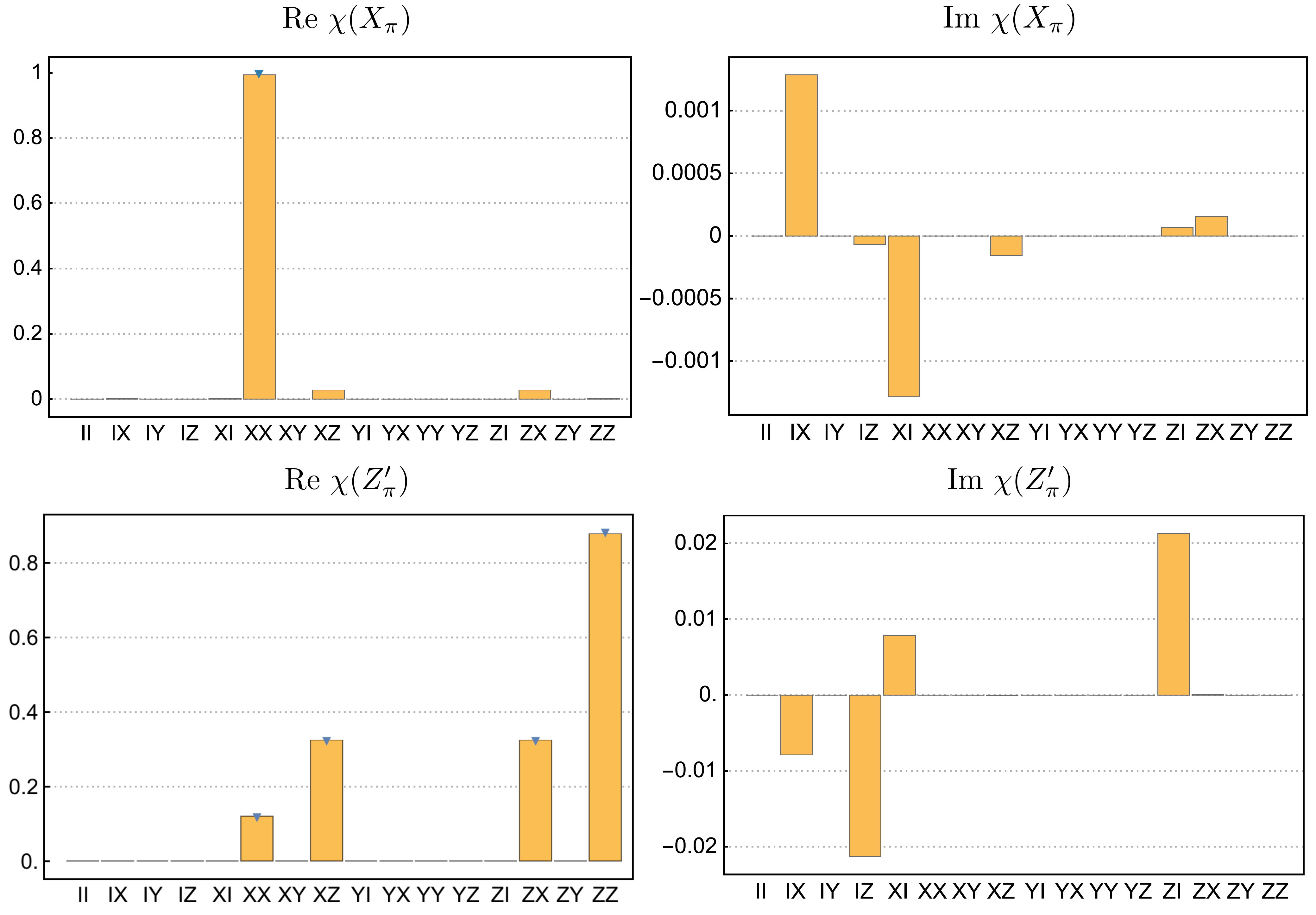}
\caption{(Color online)
The real and imaginary parts of the $\chi$ matrix elements for the fully optimized $X_\pi$ and $Z'_\pi$ gates, obtained from Eq.~\eqref{chiEq}.  Triangles indicate nonzero target values of the $\chi$ matrix elements.}
\label{chiFig}
\end{center}
\end{figure*}

{Finally, we address the question of whether misalignment of the magnetic fields and field gradients can degrade the estimated gate fidelities.
Recall that our proposed experimental geometry requires that ${\Delta {\bf B}\cdot {\bf B_\text{avg}}=0}$, as discussed in Sec.~\ref{sec:overview}.
Misalignment of the external field, and therefore $\bf B_\text{avg}$, causes leakage into the state $\ket{T_0}$, which has a similar effect as the Overhauser noise term $\Delta h_z$.
Using Eq.~(\ref{Pleak}) as a guide, we can estimate the leakage probability $P_0$ from Rabi's formula by replacing the matrix element ${\D B_x/\sqrt{2}\to\D B_z}$, and the energy splitting $2B_z\to B_z$.
We then obtain an estimate for the ratio of the leakage envelopes:   $P_0/P_+$$\simeq$$8(\D B_z/\D B_x)^2$.  
Considering a maximum 10\% misalignment of the magnetic fields (i.e., ${\D B_x/\D B_z<0.1}$), we see that ${P_0< 0.08\, P_+}$.  
Moreover, from Fig.~4, we see that ${P_+<0.013}$ at the optimal working, yielding an estimate of ${P_0<10^{-3}}$, which is negligible to the accuracy of results reported in the main text.}

\section{Single-qubit process tomography\label{chi}}
{
Here, we outline our calculation of the single-qubit process matrix in Eq.~(\ref{eq:tomo}), following Ref.~[\onlinecite{NielsenBook}].
We simulate the evolution starting from each of four initial states, which are expressed in conventional notation as $|Z\>=|S\>$, $|{\rm -}Z\>=|T_-\>$, $|X\>=(|S\>+|T_-\>)/\sqrt{2}$, and $|Y\>=(|S\>+i|T_-\>)/\sqrt{2}$, with reference to the qubit basis states $\ket{S}$ and $\ket{T_-}$.
Each initial state evolved in time using the full 5D Hamiltonian machinery described in Sec.~\ref{gf}, yielding results that include both noise and leakage.
We then perform a Gaussian average over the quasistatic noise variables, also described in Sec.~\ref{gf}, yielding the final state density matrices
\begin{align}
\rho_1'&=\mcal{E}(|{\rm -} Z\rangle\langle{\rm -}Z|) , \\
\rho_4'&=\mcal{E}(|Z\rangle\langle Z|) , \nn
\rho_2'&=\mcal{E}(|X\rangle\langle X|)-i\mcal{E}(|Y\rangle\langle Y|)-(1-i)(\rho_1'+\rho_4')/2 , \nn
\rho_3'&=\mcal{E}(|X\rangle\langle X|)+i\mcal{E}(|Y\rangle\langle Y|)-(1+i)(\rho_1'+\rho_4')/2 . \notag
\end{align}
The matrices $\rho_1', .. ,\rho_4'$ have dimension $5\times 5$, yielding $\chi$ matrices that also have a high dimension.
However, when we compute gate fidelities as described in Sec.~\ref{gf}, the product $\chi(\mcal{E})\chi(\mcal{E}_0)$ effectively collapses onto the 2D qubit subspace.
To make use of the conventional formulation of process tomography~\cite{NielsenBook}, we therefore project the density matrices $\rho_1', .. ,\rho_4'$ onto the 2D qubit subspace from the outset.
We then proceed to construct the process matrices in the usual way~\cite{NielsenBook}, as
\ben
\chi=\Lm\begin{pmatrix}\rho_1'&\rho_2'\\\rho_3'&\rho_4'\end{pmatrix}\Lm\,,\quad
\Lm=\onehalf\begin{pmatrix}I&\hat{\tau}_x\\\hat{\tau}_x&-I\end{pmatrix}\,.
\label{chiEq}
\een
The real and imaginary parts of the $\chi$ matrix elements formed in this way are plotted in Fig.~\ref{chiFig} for $X_\pi$ and $Z'_\pi$ gates.}

\section{Qubit dephasing in the $S$-$T_-$  subspace\label{dephase}}
In this Appendix, we derive explicit formulas for dephasing due to quasistatic fluctuations of the Overhauser fields and the detuning parameter, focusing on the $X$ rotations \footnote{
{Here, we neglect $T_1$ relaxation processes, since experimental measurements of singlet-triplet lifetimes are on the order of ms at zero magnetic field (and longer in applied fields~\cite{{Prance:2012p046808}}), and are therefore much longer than any $T_2^*$ time scales studied here.}
}.
For simplicity, we limit our discussion to the 2D qubit subspace, where, up to an overall shift in the energy, the qubit Hamiltonian is then given by $H^{(ST_-)}$=$(\mbb+\de\mbb)\cdot\hat{\bm{\tau}}/2$, where 
\ben
\mbb(\e)=-g\mu_B\left[{\D B_x\over\sqrt{2}}\hatbf{x}+\left(B_z-{J(\e)\over g\mu_B}\right)\hatbf{z}\right]
\label{b}
\een
is the constant effective field, and 
\ben
\de\mbb=-g\mu_B\left[\frac{\D h_x}{\sqrt{2}}\hatbf{x}+{{\D h_y}\over\sqrt{2}}\hatbf{y}
+\left(h_z-{\de J\over g\mu_B}\right)\hatbf{z}\right]
\label{db0}
\een
is the fluctuating component due to the noise terms.  
Here, we keep only the leading order terms in the effective Hamiltonian, and we take $\cos\eta\simeq 1$, which is a very good approximation near the qubit working points, $\epsilon_X$ and $\epsilon_{Z'}$. 
In Eq.~\eqref{db0}, we have defined $\de J$=$J(\e+\de\e)$$-$$J(\epsilon)\simeq(\p J/\p\e)\de\e$ as the fluctuation of the exchange energy, which arises from fluctuations of the detuning parameter, $\delta \epsilon$.
The characteristic variance of the exchange fluctuations is related to the detuning variance as $\s_J=(\p J/\p\e)\s_\e$.  
Here, $\hat{\bs{\tau}}$ denotes the Pauli matrices spanning the $S$-$T_-$ subspace.  (For example, $\hat{\tau}_z\equiv |S\>\<S|-|T_-\>\<T_-|$.)

\subsection{Pure dephasing rates\label{T2eps}}
We now obtain an expression for inhomogeneous broadening ($T_2^*$) in the $S$-$T_-$ subspace.
The energy splitting between the energy eigenstates $\ket{0}$ and $\ket{1}$ of the qubit Hamiltonian, Eq.~\eqref{H2D}, is given by $E_{01}$=$|\mbb|$, and its fluctuation, up to quadratic order in the noise terms, is given by
\ben
\de E_{01}\simeq \de b_\|+\frac{|\de\mbb_\perp|^2}{2E_{01}}\,,
\label{dE}
\een
where $\de b_\|$=$\de\mbb\cdot\hatbf{b}$ and $\de\mbb_\perp$=$\de\mbb-\de b_\|\hatbf{b}$ are the components of the noise field longitudinal and transverse to $\hatbf{b}$=$\mbb/|\mbb|$, respectively. 

The expansion in Eq.~(\ref{dE}) requires that $\de E_{01}/E_{01}$$\ll$$1$, which could potentially be violated in some cases, particularly at the working point $\epsilon_X$, where the energy splitting, $E_{01}=g\mu_B\Delta B_x/\sqrt{2}$, is relatively small.
To check this, we first identify the individual fluctuation contributions to $\delta E_{01}$ in Eqs.~(\ref{b}) and (\ref{db0}).
[For example, $(\partial E_{01}/\partial \epsilon)\delta \epsilon$ is the contribution from detuning noise.]
For the detuning fluctuations, we note that $\partial E_{01}/\partial J=0$ when $\epsilon=\epsilon_X$, while $\partial J/\partial \epsilon \simeq 0$ when $\epsilon=\epsilon_{Z'}$.
Moreover, we have previously noted that $\sigma_\epsilon,\sigma_h\simeq 3$~neV, while $\sigma_J\simeq 7$~neV, and $g\mu_B(\Delta B_x,B_z)\simeq (0.25,0.75)$~$\mu$eV at the optimal working point.
With this information, it is easy to show that Eq.~(\ref{dE}) is satisfied.


The pure dephasing times are computed by averaging the dynamical phase difference between states $|0\>$ and $|1\>$ over the different noise variables~\cite{merkulov:2002p205309}:
\begin{equation}
\overline{e^{i(\de E_{01}t)/\hbar }} =\overline{e^{i\de b_\|t/\hbar}}\overline{e^{i{\de\mbb^2_\perp}/{2E_{01}}}}, \label{phase} 
\end{equation}
where the overbar denotes the noise average.  
If we assume Gaussian distributions for the noise variables $\de b_i$, with variances $\s_i$, the average of a generic function ${g(\de\mbb)}$ is given by
\[\overline{g(\de\mbb)}=\prod_i\int\frac{d(\de b_i)}{\sqrt{2\pi}\s_i}{g(\de\mbb)} e^{-\de b_i^2/2\s_i^2}\,,\]
where $\overline{\de b_i}$=$0$, $\overline{\de b_i\de b_j}$=$\de_{ij}\s_i^2$, and $i$=$x,y,z$.  
We first note that the longitudinal and transverse noise integrals in Eq.~\eqref{phase} are separable because noises in orthogonal directions are uncorrelated, so that
\begin{equation}
\overline{e^{i(\de E_{01}t)/\hbar }}=e^{-(t/T_2^*(\e))^2}W_\perp(t) ,
\end{equation}
where
\begin{gather}
e^{-(t/T_2^*(\e))^2} \equiv \overline{e^{i\de b_\|t/\hbar}} , \label{phaset} \\
W_\perp(t) \equiv \overline{e^{i{\de\mbb^2_\perp}/{2E_{01}}}} ,
\label{phaseW}
\end{gather}

We then consider the leading, longitudinal contributions to the noise, which gives rise to ${T_2^*(\e)}$.  
Evaluating the noise integral in Eq.~\eqref{phaset}, we obtain
\begin{widetext}
\begin{equation}
\frac{\sqrt{2}\hbar}{T_2^*(\e)}=\sqrt{\left(\sigma_J(\e)  \frac{\partial E_{01}}{\partial J }\right)^2+\sigma _h^2
\left[\pfrac{\partial E_{01}}{\partial {\Delta B}_x}^2+\pfrac{\partial E_{01}}{\partial { B}_z}^2\right]} 
=\sqrt{\frac{(\sigma _h^2+\sigma _J(\e)^2) (J(\e)/g\mu_B-B_z)^2+\sigma _h^2\Delta B_x^2/2}{(J(\e)/g\mu_B-B_z)^2+\Delta B_x^2/2}}\,.
\label{T2}
\end{equation}
\end{widetext}
At the sweet spot $\e_X$, where $J=g\mu_BB_z$, Eq.~(\ref{T2}) reduces to the simple form $T_2^*(\epsilon_X)=\sqrt{2}\hbar/\sigma_h$.  
Here, we note that $\epsilon_X$ is also a sweet spot for the transverse fluctuations of the Overhauser fields, $h_z$ and $\D h_y$, because  $\partial E_{01}/\partial B_z$=0 and $\partial E_{01}/\partial(\D B_y)$=0.
For $Z'$ rotations, which occur in the far-detuned region where $J\simeq0$ and $\sigma_J\simeq 0$, Eq.~\eqref{T2} also predicts that $T_2^*(\epsilon_{Z'})\simeq \sqrt{2}\hbar/\sigma_h$.  
It is to be expected that the Overhauser field fluctuations determine the dephasing times for both rotation axes because they are both driven by magnetic fields.
When optimal tuning parameters are used, the secondary, transverse contributions to the inhomogeneous broadening, which yield the correction term $W_\perp(t)$, play a prominent role only when the longitudinal Overhauser field fluctuations are suppressed.
We analyze the latter situation in Appendix~\ref{nuclear}.  

{Finally, we note that pure dephasing also occurs due the electron-phonon coupling, which causes energy fluctuations between singlet and triplet states not included in Eq.~(\ref{db0}).
The associated dephasing rate was previously estimated for a double quantum dot to be $\sim10$ kHz ~\cite{Hu:2011p165322}, negligible compared to dephasing rate from Overhauser fluctuations, which is of order MHz.
Because the dominate contribution to this dephasing process comes from the dipole charge distribution of the $|S(0,2)\>$  state, the associated dephasing rate goes as $\sin^2\eta$, the probability for occupation of $\ket{S(0,2)}$.  
Therefore, at the working point $\e_X$, this dephasing rate is suppressed by ${\sin\eta^2\simeq(J/t_c)^2=(g\mu_B B_z/t_c)^2=10^{-3}}$  as reported in Sec.~\ref{ddh}, and even further suppressed at the working point $\epsilon_{Z'}$, where $J\simeq 0$.}


\subsection{Comparison with the $S$-$T_0$ qubit }\label{T2compare}

A key advantage of the $S$-$T_-$ qubit is the presence of a detuning sweet spot at $\e=\e_X$, which protects the $X$ rotations from detuning noise.
The $S$-$T_0$ qubit has a very similar energy level diagram, with similar operating points in the far-detuned regime (the $\Delta B_z$ gate), and the intermediate tuning regime (the $J$ gate).
However, the $J$ gate does not operate at a sweet spot.
In this Appendix, we investigate some consequences of this distinction.

A common figure of merit for qubit rotations is the $Q$-factor, defined by $Q=T_2^*/\tau_{2\pi}$, which describes the number of coherent oscillations that can be observed in the presence of an uncorrelated noise source.
For the $S$-$T_-$ qubit, we have already shown that $T_2^*\simeq\sqrt{2}\hbar/\sigma_h$ at the two operating points, $\epsilon_X$ and $\epsilon_{Z'}$, while the corresponding $\pi$-gate times are given by $\tau_{X_\pi}=h/\sqrt{2}g\mu_B\Delta B_X$ and $\tau_{Z'_\pi}=h/2g\mu_BB_z$.
The resulting $Q$-factors for the $S$-$T_-$ qubit are given by $g\mu_B\Delta B_x/2\pi\sigma_h$ and $g\mu_BB_z/\sqrt{2}\pi\sigma_h$ for $X$ and $Z'$ rotations, respectively.

For $S$-$T_0$ qubits, the effective field acting on the Bloch sphere is $\mbb=J\hatbf{z}+g\mu_B\D B_z\hatbf{x}$, yielding the dephasing rate~\cite{dial:2013p146804}
\[\frac{\sqrt{2}\hbar}{T_2^{*(ST_0)}}=\sqrt{\frac{(\s_J(\e)J)^2+2(\s_h g\mu_B\D B_z)^2}{J^2+(g\mu_B\D B_z)^2}}\,, \]
and the rotation frequency $\hbar \omega=\sqrt{J^2+(g\mu_B\D B_z)^2}$.
In the asymptotic regimes dominated by magnetic field or exchange couplings, the $Q$-factors are given by 
$g\mu_B\Delta B_z/2\pi\sigma_h$  for the $\Delta B_z$ gate, and $J/\sqrt{2}\pi\sigma_J$ for the $J$ gate.
We see that the $J$ gate has a different scaling behavior than the other gates we have considered so far.
Specifically, it differs from the corresponding gate in the $S$-$T_-$ qubit (the $X$ gate) because $\epsilon_X$ occurs at a sweet spot.
In practice, it is found the $J$ gate limits the qubit fidelity and that the ratio $J/\sigma_J$ cannot be improved by optimizing the detuning, due to the physical constraints on the device~\cite{dial:2013p146804}.
On the other hand, the ratio $\Delta B/\sigma_h$ can be improved through device design:
$\Delta B$ can be enhanced by engineering the micromagnet, while $\sigma_h$ can be suppressed by using a spin-0 material such as $^{28}$Si, or by narrowing the magnetic noise distribution~\cite{Bluhm:2010p216803}.
Hence, there are opportunities for improving the limiting gate fidelities in $S$-$T_-$ qubits that are not available for $S$-$T_0$ qubits.

\subsection{Crossover to dephasing dominated by charge noise\label{nuclear}}

\begin{figure}[t]
\begin{center}
\includegraphics[width=2.2in]{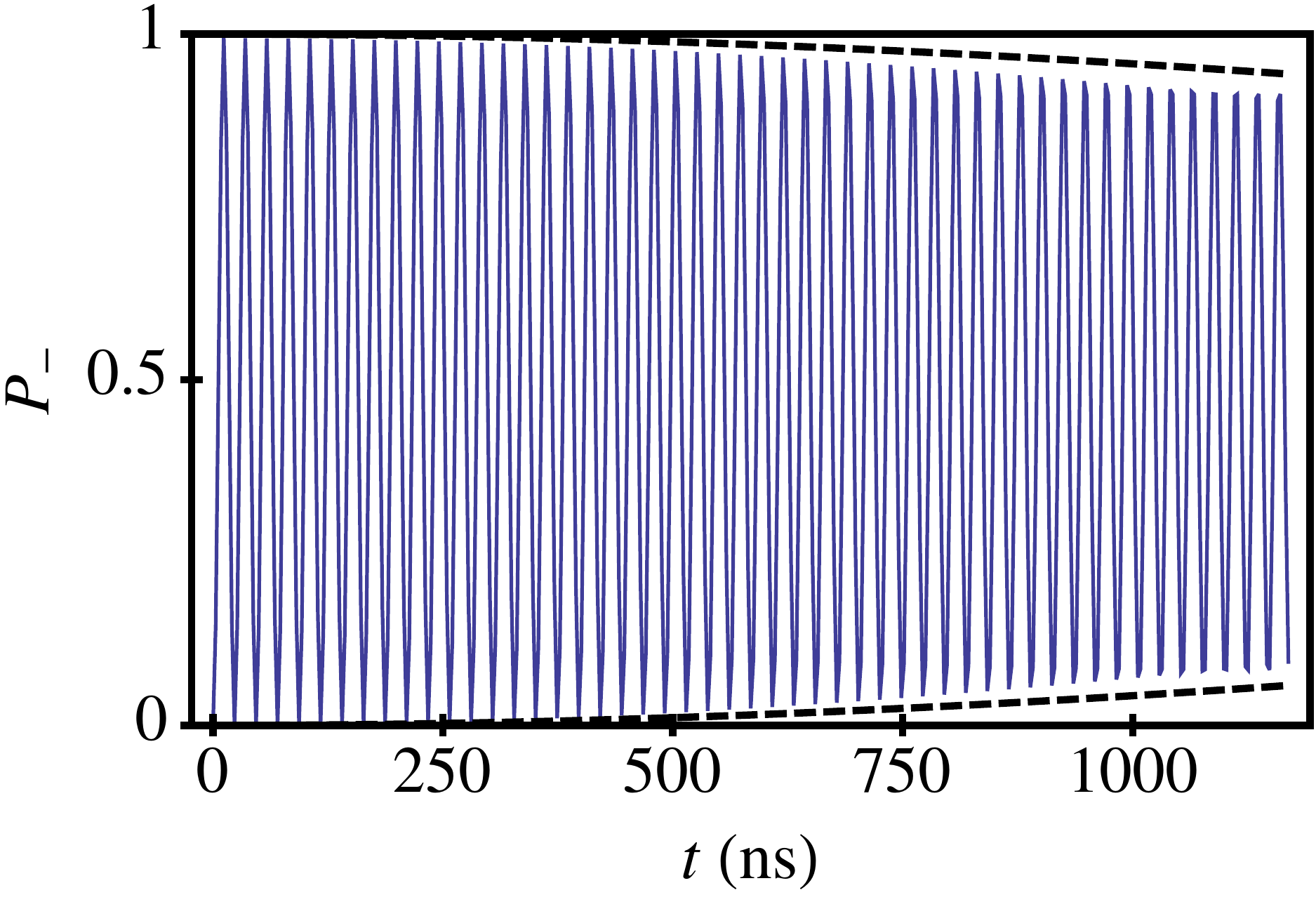}
\caption{(Color online)
The numerically computed probability $P_-$ of occupying the state $\ket{T_-}$ as a function of time $t$ during $X$ rotations in the absence of Overhauser field fluctuations, assuming the initial state $\ket{S}$.  
The dephasing envelope (dashed lines) is the Gaussian decay given by Eq.~\eqref{eq:Xenvelope}, \eqref{T2perp}, and \eqref{T2total}, setting $\s_h=0$, with $T_{2\perp}^*(X)=4.7~\mu$s.}
\label{Pm}
\end{center}
\end{figure}

In Appendix~\ref{T2eps}, we showed that the dephasing rate for $X$ rotations is usually determined by the linear-order, longitudinal noise term in Eq.~(\ref{dE}), corresponding to Overhauser fluctuations.
However, if the Overhauser fluctuations are removed by isotopic purification or by dynamic polarization, then detuning fluctuations will predominate.  
In this Appendix, we estimate when this crossover should occur.

As noted previously, the contribution to $T_2^*$ in Eq.~(\ref{T2}) from detuning noise vanishes at the operating point $\epsilon=\epsilon_X$.
However, it does not vanish for the quadratic noise term in Eq.~(\ref{dE}), and it appears in the decay envelope $W_\perp(t)$ in Eq.~\eqref{phaseW}.
The Gaussian integral in Eq.~\eqref{phaseW} can be solved exactly, yielding
\begin{equation}
W_\perp(t)=\prod_\al\frac{1}{ \sqrt{1-i\s_\al^2t/\hbar b}} , \label{eq:Wtemp}
\end{equation}
where the product is over the two transverse noise sources $\sigma_J$ and $\sigma_h$, and $b$=$|\mbb|$ is the energy splitting between the two qubit states.
At the working point $\epsilon_X$, we take $b={g\mu_B|\D B_x|/\sqrt{2}}$.
The short time limit $\sigma_\alpha^2\tau_X/\hbar b\ll1$ is easily satisfied for an $X_\pi$ gate, whose gate time is given by $\tau_{X_\pi}=h/\sqrt{2}g\mu_B\D B_x$. 
In this same limit, we can rewrite Eq.~(\ref{eq:Wtemp}) as 
\begin{equation}
W_\perp(t)
\simeq e^{i\phi}  e^{-\left(t/T_{2\perp}^*(X)\right)^2} \quad\quad (\s_\al^2t/{2\hbar b}\ll1) , 
\label{W}
\end{equation}
where $\phi$ is an irrelevant phase, and the transverse dephasing time is given by
\ben
T_{2\perp}^*(X)={{2\hbar b}\over{\s_J^2+\s_h^2}}\,.
\label{T2perp}
\een
If we denote the longitudinal dephasing time obtained in Appendix~\ref{T2eps} as ${T_{2\|}^*(X)}$=$\sqrt{2}\hbar/\s_h$, then the combined, short-time decay function at $\epsilon_X$ takes the form 
\begin{equation}
\exp(-t^2/[\til{T}_2^*(X)]^2) ,\label{eq:Xenvelope}
\end{equation}
where
\begin{equation}
[\til{T}_{2}^*(X)]^{-2}= [{T}_{2\perp}^*(X)]^{-2}+ [{T}_{2\|}^*(X)]^{-2} \,.\label{T2total}
\end{equation}
Here, the shortest dephasing time naturally dominates the decay.

We now determine the crossover from nuclear to detuning noise-dominated decay, which can be implemented by isotopically purifying the sample with $^{28}$Si.
The ratio of the longitudinal and transverse dephasing times is given by
\[\frac{T_{2\|}^*(X)}{T_{2\perp}^*(X)}=\frac{\s_J^2}{\sqrt{2}b\s_h}\left[1+\pfrac{\s_h}{\s_J}^2\right]\,. \]
Solving for the crossover point which occurs when the ratio is equal to $1$, we find $\s_h\simeq\s_J^2/\sqrt{2}b=0.2$~neV.
This happens when the abundance of $^{29}$Si is reduced from $\sim$$5\%$ (natural abundance) to $\sim$$0.5\%$.   
At this level of purification, the total dephasing time is given by ${\til{T}_2^*(X)=3.3~\mu}$s.  
In the limit of pure $^{28}$Si,  we find a dephasing time of $T_{2\perp}^*(X)=4.7~\mu$s, which is due entirely to detuning noise.  
We confirm this limiting behavior in Fig.~\ref{Pm}, which shows the numerically computed $P_-$ probability for $X$ rotations.
The short-time dephasing envelope, shown as a dashed line, is given by Eq.~(\ref{W}), where we set $\s_h$=$0$.  
Note that we have not re-optimized with respect to magnetic fields in this figure, which would yield further improvements in the coherence time.

\subsection{Analytic solution in the $S$-$T_-$ qubit subspace\label{qEOM}}
In this Appendix, we obtain approximate solutions for the equations of motion of the $S$-$T_-$ qubit.
We closely follow Ref.~[\onlinecite{merkulov:2002p205309}], which examines the relaxation dynamics of a spin in an external magnetic field, in the presence of quasistatic Overhauser field fluctuations.
We use these results to obtain Eq.~(\ref{Pma}) of the main text.

We consider the expectation value of the logical qubit pseudospin $\mbs$=$\<\psi|\hat{\bm{\tau}}|\psi\>/2$, where $|\psi\>=c_S|S\>+c_-|T_-\>$ is the state vector in the qubit subspace. 
The dynamics of $\ket{\psi}$ are governed by the 2D Hamiltonian, Eq.~\eqref{H2D}. 
As noted above, near the sweet spot $\e_X$, the relevant noise sources all have variances $\s_\al$ of order neV, so that $\s_\al/\D B_x$$\sim$$1\%$.  
Following Ref.~[\onlinecite{merkulov:2002p205309}], we can therefore obtain solutions for $\bf s$ to leading order in $\s_\al/\D B_x$.
For an initial state $|S\>$=$|Z\>$, we find that
\begin{align}
\overline{\mbs(t)}&={e^{-\left(t/T_2^*(X)\right)^2}\over2}\label{s}\\ \times
&\left[\left(\cos(\w_X t)+2\de_\perp^2\sin^2 (\w_X t/2)\right)\hatbf{z}-\sin(\w_X t)\hatbf{y}\right]\,,\nonumber
\end{align}
where, as usual, $\w_X=b/\hbar$, $b$=${g\mu_B|\D B_x|/\sqrt{2}}$,  $T_2^*(X)=\sqrt{2}\hbar/\sigma_h$, and
\ben
\de_\perp^2=\frac{\s_J^2+\s_{h}^2}{b^2}\, .
\label{T2x}
\een
Here, $\s_J=\s_\e(\p J/\p\e)$ is evaluated at $\epsilon=\epsilon_X$.  
Equation~(\ref{Pma}) in the main text is obtained by evaluating Eq.~(\ref{s}) at the gate time $\tau_{X_\pi}=\pi/\omega_{X}$, using the definition $P_-'(\tau_X)$$=$${1/2}-\overline{s_z}(\tau_X)$. 


\section{Limiting behavior when $\epsilon\rightarrow -\infty$ \label{qEOM1}}

In this Appendix, we explain why the second sweet spot for the $S$-$T_-$ qubit, which occurs in the limit $\epsilon\rightarrow -\infty$, is not an optimal working point.
Our investigation is simplified by the fact that $J\rightarrow 0$, making exact analytical solutions possible.
We find that although dephasing effects due to charge noise are suppressed, leakage is enhanced. 

The full 5D Hamiltonian in Eq.~(\ref{H}) simplifies in the limit $J\rightarrow 0$.  
Moreover, the $(0,2)$ charge state is split off by a large energy, and only the four $(1,1)$ charge states are accessible.
Equation~(\ref{H}) can then be rewritten as a pure spin Hamiltonian:
\ben
H=\frac{1}{2}g\mu_B\sum_{i=L,R}(\mbB_i+\mbh_i)\cdot \bm{\s}_{i} \,,
\label{H1}
\een
where $\bm{\s}_{L}$ and $\bm{\s}_{R}$ denote the Pauli spin operators on the left and right dots, respectively.  
(Note that these operators differ from the $\hat{\bm{\tau}}$ operators previously defined for the logical qubit states.)
The resulting time evolution operator is given by
\begin{widetext}
\begin{align}
U(t)&=\exp(i\w_Lt\hatbf{n}_L\cdot\bm{\s}_L)\otimes \exp(i\w_Rt\hatbf{n}_R\cdot\bm{\s}_L)
\label{eq:U} \\ \nonumber
&=\cos(\w_Lt)\cos(\w_Rt) +i\cos(\w_Lt)\sin(\w_Rt)\sum_{i=L,R} \hatbf{n}_i\cdot\bm{\s}_i-\sin(\w_Lt)\sin(\w_Rt) (\hatbf{n}_L\cdot\bm{\s}_L)\otimes(\hatbf{n}_R\cdot\bm{\s}_R)\, ,
\end{align}
where we define
\[ \w_i={g\mu_B|\mbB_i+\mbh_i|\over2\hbar} \quad \text{and}\quad 
 \hatbf{n}_i=\frac{\mbB_i+\mbh_i}{|\mbB_i+\mbh_i|},  \quad (i=L,R)\, . \]

Equations~(\ref{H1}) and (\ref{eq:U}) describe the precession of individual spins about their local magnetic fields.  
The system dynamics is similar to that of the $Z'$ rotation, which occurs at finite $\epsilon$, and we can estimate the rotation fidelity in a similar way.
We consider the time evolution of $U(t)$ on the initial state  ${|X\>=(|S\>+|T_-\>)/\sqrt{2}}$.
The relevant transition amplitude is readily computed from
\begin{equation}  
{\<{\rm-}X|U(t)|X\>}={1\over2}(\<S|U|S\>-\<T_-|U|T_-\>+2i\, {\rm Im}[\<S|U|T_-\>]) \,,
\label{sol}
\end{equation}
where
\begin{align*}  
\<S|U(t)|S\>&=\cos\w_Lt\cos\w_Rt+\sin\w_Lt\sin\w_Rt(\hatbf{n}_L\cdot\hatbf{n}_R)\,,\nn
\<S|U(t)|T_-\>&={1\over\sqrt{2}} \left[\sin\w_Lt\sin\w_Rt(n_L^z n_R^+-n_L^+n_R^z)-i\cos\w_Lt\sin\w_Rt (n_L^+-n_R^+)\right]\,,\nn
\<T_-|U(t)|T_-\>&=\cos\w_Lt\cos\w_Rt  -i\cos\w_Lt\sin\w_Rt (n_R^z+n_L^z)-\sin\w_Lt\sin\w_Rt(n_L^z n_R^z)\,,
\end{align*}
\end{widetext}
and $n_i^\pm=n_i^x\pm in_i^y$ ($i=L,R$).
We compute the probability 
$P(-X)$=$\overline{|\<{\rm-}X|U(t)|X\>|^2}$ of occupying the state $|{\rm-}X\>$ as a function of time.  
Here, the overbar indicates an average of the Overhauser field fluctuations, taken over all six directional field components of $\mbh_{L}$ and $\mbh_{R}$.   
The Overhauser averages are obtained numerically, assuming Gaussian distributions for the fluctuating variables.

\begin{figure}[t]
\includegraphics[width=2.2in]{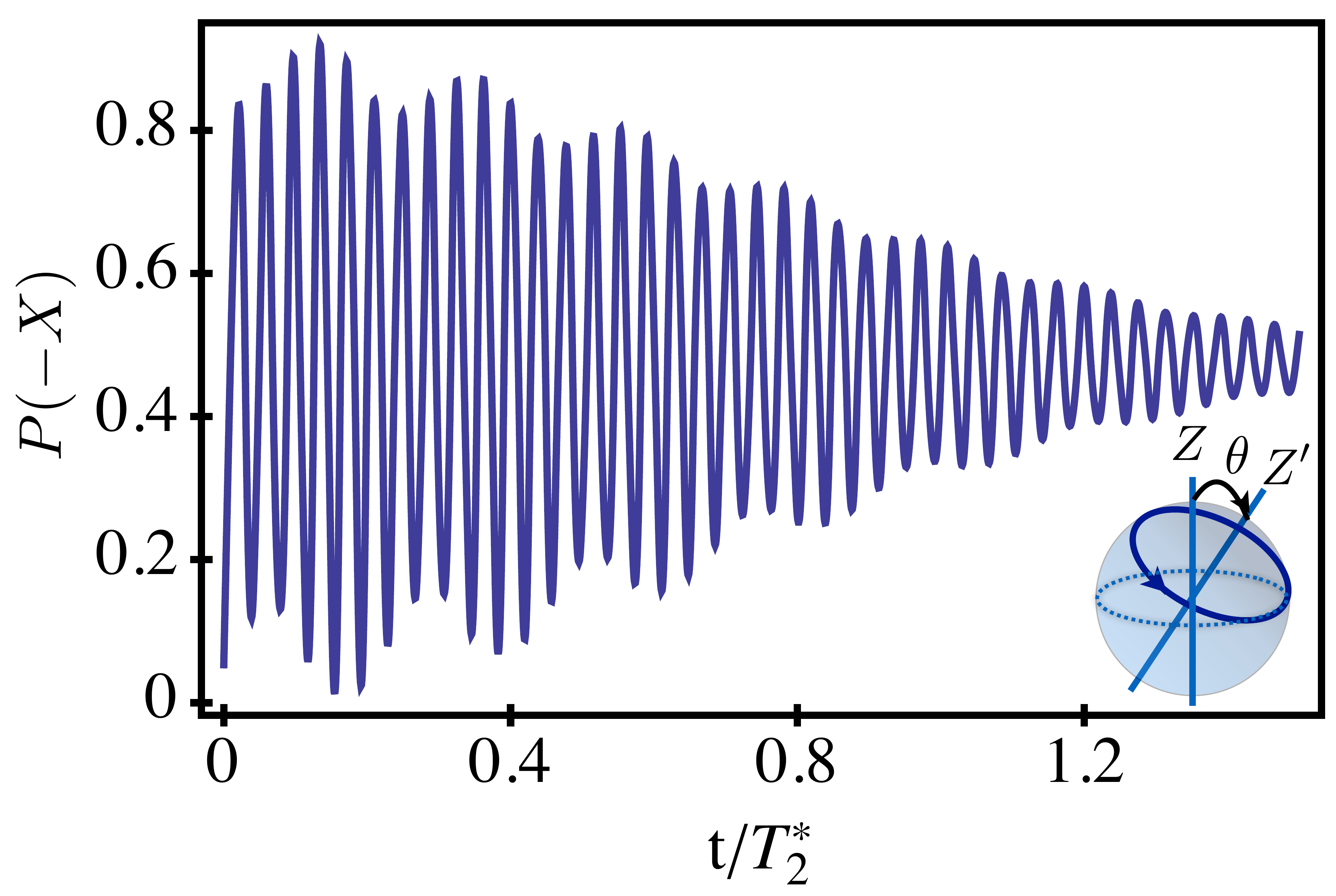}
\caption{(Color online)
$Z'$ rotations in the asymptotic regime $\epsilon\rightarrow -\infty$, obtained from analytic solutions for the qubit dynamics arising from Eq.~(\ref{H1}).
Here, we plot the probability $P(-X)$ of being in final state $\ket{-X}$ as a function of time $t$, for initial state $\ket{X}$ and the optimal magnetic fields $(\D B_x,B_z)$=$(0.25,0.75)~\mu$eV.
The analytic solutions are numerically averaged over the Overhauser field fluctuations.}
\label{FID}
\end{figure}

The resulting probability $P(-X)$ is plotted in Fig.~\ref{FID} in time units of $T_2^*$=$\sqrt{2}\hbar/\s_h$.
Here,  we assume optimal magnetic field values and $\D B_z$=$0$ as in the main text, so that leakage to $T_0$ vanishes, though we note that our solution is valid for $\D B_z\neq0$.
It is instructive to compare these results with the analogous $Z'$ oscillations obtained when $J>0$, which are plotted in Fig.~\ref{fig:Zrot}(b) of the main text.  
The most striking feature of the ${J=0}$ oscillations is the modulation of the envelope, caused by leakage to $\ket{T_+}$.  
Fig~1 (b) shows why leakage is enhanced in the $\epsilon\rightarrow -\infty$ limit: the energy gap between $\ket{S}$ and $\ket{T_+}$ that suppresses leakage is at the minimum $g\mu_BB_z$, compared to $g\mu_BB_z+J(\e)$ at finite detuning.  The additional suppression with $J>0$ makes $\epsilon=\epsilon_{Z'}$ a preferable working point.

Finally, we note that Eqs.~(\ref{H1}) and (\ref{eq:U}) represent exact solutions in the limit $\epsilon\rightarrow -\infty$.
These expressions do not contain any reference to $\epsilon$, and the resulting dynamics is unaffected by detuning noise.
By performing an average over quasistatic Overhauser field fluctuations in the dynamics, we can directly probe the magnetic variance $\s_h$ by comparing to experimental data, as was done for the $S$-$T_0$ qubit in Refs.~[\onlinecite{Petta:2005p2180}] and~[\onlinecite{Taylor:2005p177}].

\bibliography{citations}
\end{document}